\documentclass[conference]{IEEEtran}
\usepackage{color,graphicx}
\usepackage{url}
\usepackage{amsmath}
\usepackage{multirow}
\usepackage{caption}
\usepackage{algorithmic}
\usepackage{algorithm}
\usepackage[algo2e]{algorithm2e} 
\usepackage{epstopdf}
\usepackage{epsfig}
\definecolor{Gray}{gray}{0.9}
\usepackage[table]{xcolor}
\usepackage{subcaption}
\usepackage{tikz}

\makeatletter

\ifCLASSINFOpdf
\else
\fi


\begin{document}
\title{Machine Learning enabled Spectrum Sharing in Dense LTE-U/Wi-Fi Coexistence Scenarios}
\author{Adam Dziedzic$^\dag\text{*}$, Vanlin Sathya$^\dag\text{*}$, Muhammad Iqbal Rochman$^\dag$, Monisha Ghosh$^\dag$, and Sanjay Krishnan$^\dag$
\IEEEauthorblockN{}
\IEEEauthorblockA{$^\dag$University of Chicago, Illinois-60637, USA.\\
{Email: \{ady, vanlin, muhiqbalcr, monisha, skr\}@uchicago.edu}}
}
\maketitle

\footnote{*Equal contribution.}
\begin{abstract}
The application of Machine Learning (ML) techniques to complex engineering problems has proved to be an attractive and efficient solution. ML has been successfully applied to several practical tasks like image recognition, automating industrial operations, etc. The promise of ML techniques in solving non-linear problems influenced this work which aims to apply known ML techniques and develop new ones for wireless spectrum sharing between Wi-Fi and LTE in the unlicensed spectrum. In this work, we focus on the LTE-Unlicensed (LTE-U) specification developed by the LTE-U Forum, which uses the duty-cycle approach for fair coexistence. The specification suggests reducing the duty cycle at the LTE-U base-station (BS) when the number of co-channel \mbox{Wi-Fi} basic service sets (BSSs) increases from one to two or more. However, without decoding the \mbox{Wi-Fi} packets, detecting the number of \mbox{Wi-Fi} BSSs operating on the channel in real-time is a challenging problem. In this work, we demonstrate a novel ML-based approach which solves this problem by using energy values observed during the \mbox{LTE-U} OFF duration. It is relatively straightforward to observe only the energy values during the \mbox{LTE-U} BS OFF time compared to decoding the entire \mbox{Wi-Fi} packet, which would require a full \mbox{Wi-Fi} receiver at the LTE-U base-station. We implement and validate the proposed ML-based approach by real-time experiments and demonstrate that there exist distinct patterns between the energy distributions between one and many \mbox{Wi-Fi} AP transmissions. The proposed ML-based  approach results in a higher accuracy (close to 99\% in all cases) as compared to the existing auto-correlation (AC) and energy detection (ED) approaches.
\end{abstract}

\begin{IEEEkeywords}
LTE, Unlicensed Spectrum, Wi-Fi, Machine Learning.
\end{IEEEkeywords}

\section{Introduction}\label{sec:introduction}

\begin{figure}[htb!]
\begin{center}
\includegraphics[height=5.3cm,width=9cm]{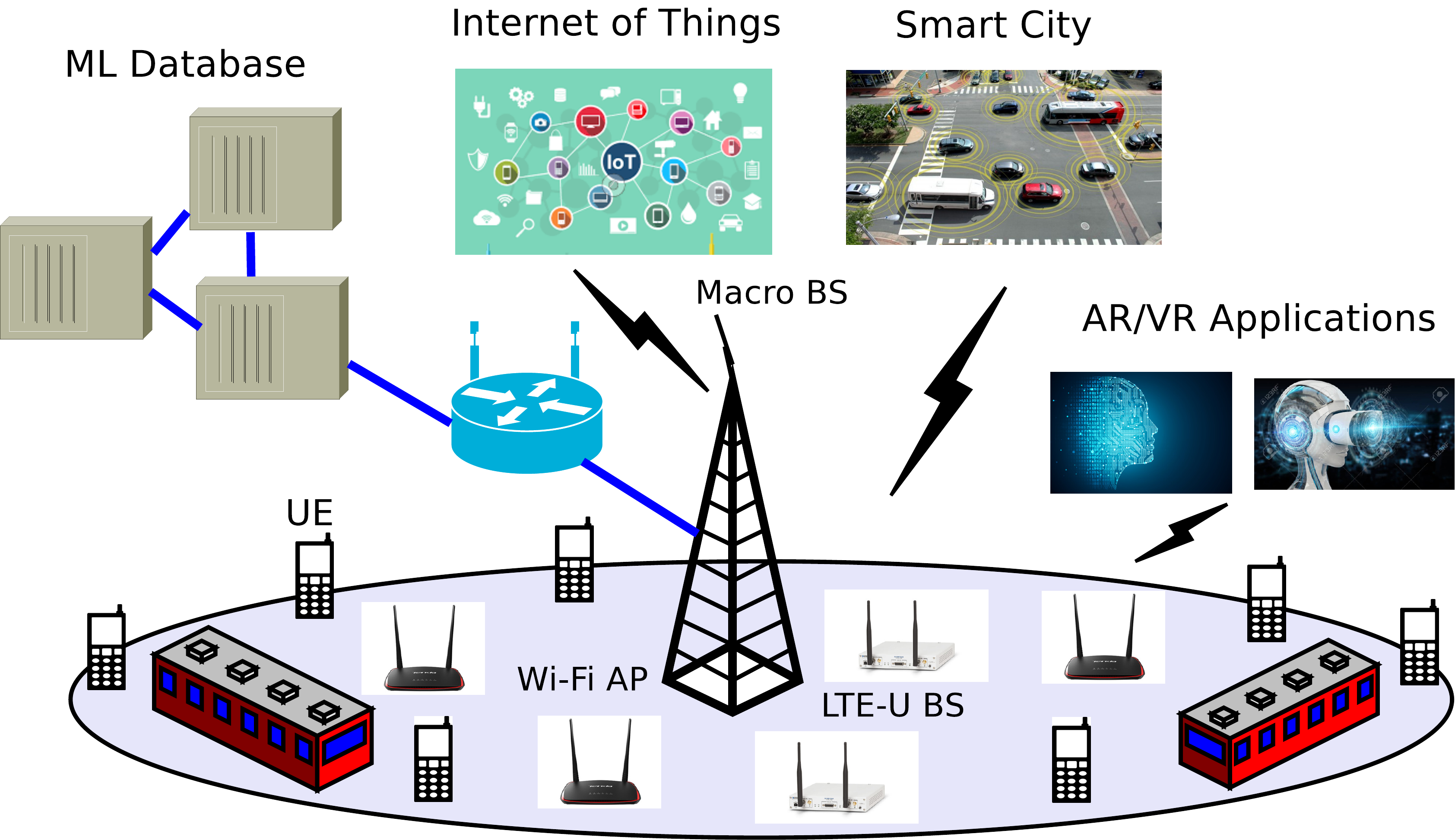}
\caption{Future Applications on Unlicensed Spectrum Band.}
\label{mle}
\end{center}
\end{figure}

The growing penetration of high-end consumer devices like smartphones and tablets running bandwidth hungry applications (e.g. mobile multimedia streaming) has led to a commensurate surge in demand for mobile data (pegged to soar up to 77 exabytes by 2022 \cite{cisco2018cisco}). An anticipated second wave will result from the emerging Augmented/Virtual Reality (AR/VR) industry \cite{al2017energy} and more broadly, the Internet-of-Things that will connect an unprecedented number of intelligent devices to next-generation (5G and beyond) mobile networks as shown in Fig.~\ref{mle}. Existing wireless networks, both cellular and Wi-Fi,  must therefore greatly expand their aggregate {\em network} capacity to meet this challenge. This is being achieved by a combination of approaches including use of multi-input, multi-output (MIMO) techniques \cite{gampala2018massive}, network densification (i.e. deploying small cells \cite{sathya2014placement}) and more efficient traffic management and radio resource allocation. 

Since  licensed spectrum is a limited and expensive resource, its optimal utilization may require spectrum sharing between multiple network operators/providers of different types -increasingly licensed-unlicensed sharing is being contemplated to enhance network spectral efficiency, beyond the more traditional unlicensed-unlicensed sharing. As the most common unlicensed incumbent, Wi-Fi is now broadly deployed in the unlicensed $5$ GHz band in North America where approximately $500$ MHz of bandwidth is available. However, these $5$ GHz unlicensed bands are also seeing increasing deployment of cellular services such as Long Term Evolution (LTE) Licensed Assisted Access (LTE-LAA). Recently, the Federal Communications Commission (FCC) sought to open up 1.2 GHz of additional spectrum for unlicensed operation in the 6 GHz band through a Notice of Proposed Rule Making (NPRM) \cite{FCC1}.  This allocation of spectrum for unlicensed operation will thus only accelerate the need for further coexistence solutions among heterogeneous systems.

However, the benefits of spectrum sharing are not devoid of challenges, the foremost being the search for effective coexistence solutions between cellular (LTE and 5G) and Wi-Fi networks whose medium access control (MAC) protocols are very different. While cellular systems employ a Time Division Multiple Access (TDMA)/Frequency Division Multiple Access (FDMA) scheduling mechanism, Wi-Fi depends on  the Carrier Sense Multiple Access with Collision Avoidance (CSMA/CA) mechanism.
The 5 GHz band being unlicensed and offering ~500 MHz of available bandwidth has prompted several key players in the cellular industry to develop the LTE-LAA specification within the Third Generation Partnership Project (3GPP). Specification differences between LTE and the incumbent Wi-Fi will lead to many issues due to the incompatibility between the two standards. Therefore, to ensure fair coexistence, certain medium access protocols have been developed as an addition to the licensed LTE standard. In addition to LTE-LAA, there also exists LTE-U which was developed by an industry consortium called the LTE-U Forum and will be the main focus of this paper.

LTE-LAA was proposed by 3GPP~\cite{3gpp,TCCN} and its working mechanism is similar to the Carrier Sense Multiple Access with Collision Avoidance (CSMA/CA) protocol used by Wi-Fi. In LTE-LAA, an LAA base station (BS) acts essentially similar to a Wi-Fi access point (AP) in terms of channel access, \textit{i.e.}, a BS needs to ensure that the channel is free before transmitting any data, otherwise it will perform an exponential back-off procedure similar to CSMA/CA in Wi-Fi. Therefore, there is no need to precisely determine the number of coexisting Wi-Fi APs, due to the channel sensing and back-off mechanism which is adaptable to varying channel occupancy. However, LTE-U which was developed by the LTE-U forum~\cite{forum}, uses a simple duty-cycling technique where the LTE-U BS will periodically switch between ON and OFF states in an interval set according to the number of Wi-Fi APs present in the channel. In the ON state, the BS transmits data as a normal LTE transmission while in the OFF state, the BS does not transmit any data but passively senses the channel for the presence of Wi-Fi. The number of sensed Wi-Fi APs is then used to properly adjust the duty cycle interval, and this process is known as Carrier Sense Adaptive Transmission (CSAT). Therefore, accurately determining the number of coexisting Wi-Fi APs is important for optimum operation of the CSAT procedure.

Existing literature addresses the LTE-U and Wi-Fi coexistence in terms of optimizing the ON and OFF duty cycle \cite{singh2018wi}, power control \cite{chaves2013lte}, hidden node problem \cite{atif2019complete}, etc. On the other hand, the LTE-U specification does not specify, and there has been relatively less work on, how a LTE-U operator should detect the number of Wi-Fi APs on the channel to adjust the duty cycle appropriately. There are a number of candidate techniques to determine the number of Wi-Fi APs as follows:

\begin{itemize}
\item \textbf{Header-Based CSAT (HD):} 
Wi-Fi APs transmit beacon packets every 102.4 ms, containing important information about the AP, such as the Basic Service Set Identification (BSSID) which is unique to each AP. This is a straightforward way to identify the Wi-Fi AP, but it adds additional complexity since the LTE-U BS would require a full Wi-Fi decoder to obtain this information from the packet.
\item \textbf{Energy-Based CSAT (ED):}
Rather than a full decoding process, it is hypothesized that sensing the energy level of the channel is enough to detect the number of Wi-Fi APs on the channel. However, it is still a challenging problem since the energy level may not correctly correlate to the number of APs under varying conditions (\textit{e.g.}, different category of traffic, large number of Wi-Fi APs, variations in transmission powers, multipath, etc).
\item \textbf{Autocorrelation-Based CSAT (AC):}
To detect the Wi-Fi signal at the LTE-U BS, one can develop an auto-correlation (AC) based detector where the LTE-U BS performs auto-correlation on the Wi-Fi preamble, without fully decoding the preamble. This is possible since all Wi-Fi preambles~\footnote{All Wi-Fi frames, even those in newer specifications like 802.11ax, begin with the legacy short training field (L-STF) symbol.} contain the legacy short training field (L-STF) and legacy long training field (L-LTF) symbols which contain multiple repeats of a known sequence. However, the AC function can only determine whether a signal is a Wi-Fi signal and cannot derive any distinct information pertaining to each APs.  
\end{itemize}
Table~\ref{table:csat} lists the different types of CSAT approaches with their own pros and cons. We studied energy detection (ED) and AC based detection of \mbox{Wi-Fi} APs in our previous work \cite{sathya2018energy}\cite{sathya2019auto}~\footnote{The latest version can be found here: http://bit.ly/2LDVWWo}, and proved that our algorithms performed reasonably well under various scenarios.

\par Of late, Machine Learning (ML) approaches are beginning to be used in wireless networks to solve problems such as agile management of network resources using real-time analytics based on data. The  advantage of ML is that it has the ability to learn useful information from input data, which can help improve network performance. ML models enable us to replace heuristics with more robust and general alternatives. In this paper, we propose observing the \mbox{Wi-Fi} AP energy values during LTE-U OFF duration and using the data to train different ML models~\cite{zhang2019deep}. We also apply the models in an online experiment to detect the number of \mbox{Wi-Fi} APs. Finally, we demonstrate significant improvement in the performance of the ML approach as compared to the ED and AC detectors.   

\begin{table}
\caption{Different Types of LTE-U CSAT.
}
\centering	
\begin{tabular}{|p{1.5cm}| p{2cm}| p{1.8cm}| p{1.8cm} |}
\hline
\bfseries
\cellcolor{Gray} CSAT Types &\bfseries \cellcolor{Gray} Method &\bfseries \cellcolor{Gray} Pros &\bfseries \cellcolor{Gray} Cons \\ 
\hline
Header Decoding (HD) & Decodes the \mbox{Wi-Fi} MAC header at the \mbox{LTE-U} BS & 100\% accurate & Additional Complexity~\cite{chai2016lte}, high cost\\
\hline
Energy Detection (ED) & Based on the change in the \textit{energy level} of the air medium & 
Low-cost, low-complexity & Low-accuracy
\cite{sathya2018energy}\\
\hline
Auto-correlation (AC) & LTE-U BS performs correlation on the \mbox{Wi-Fi} L-STF symbol in the preamble & Low-cost, low-complexity & Medium accuracy (more accurate than ED)~\cite{sathya2019auto}  \\
\hline
Machine Learning (ML) & Train the model based on energy values on the channel & Much more accurate than ED and AC methods & Requires gathering data and training models\\
\hline
\end{tabular}
\label{table:csat}
\end{table}

\begin{figure}[htb!]
\begin{center}
\includegraphics[height=6.5cm,width=9cm]{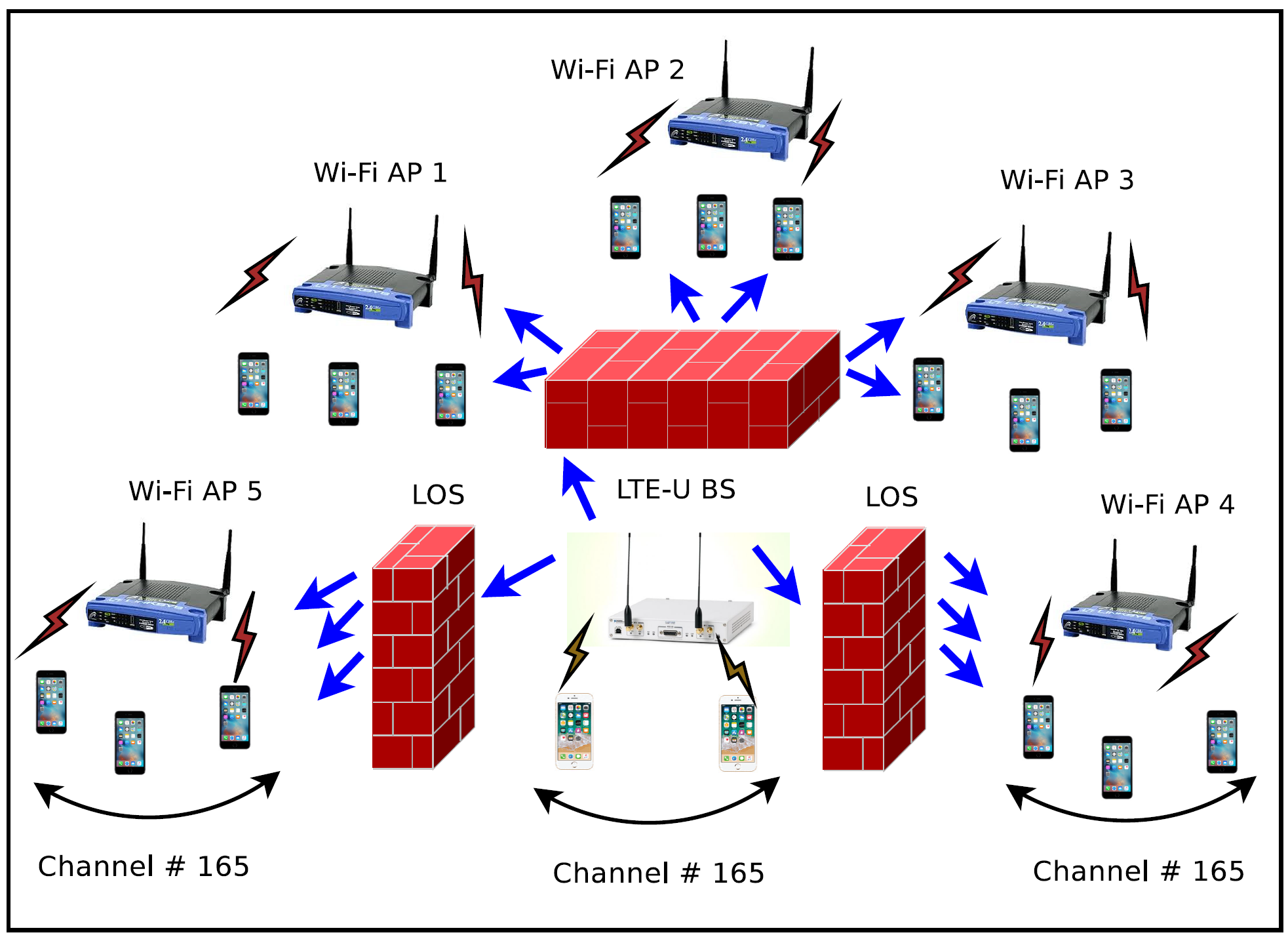}
\caption{Dense LTE \mbox{Wi-Fi} Co-existence Deployment Setup.}
\label{expp}
\end{center}
\end{figure}

\par Fig.~\ref{expp} illustrates an example of a dense LTE-U/\mbox{Wi-Fi} coexistence, where a number of \mbox{Wi-Fi} APs and one \mbox{LTE-U} BS are operating on the same channel, with multiple clients associated with each AP and BS. In such a situation, it is crucial that LTE-U reduce its duty-cycle proportional to the number of Wi-Fi APs, else with a duty-cycle of 50\% the Wi-Fi APs will be starved of air-time. As the number of Wi-Fi APs increase on the channel, it becomes increasingly important to detect the number of APs accurately at the LTE-U BS with out any co-ordination \emph{i.e.,} in a distributed manner. According to the LTE-U forum, it is expected that the \mbox{LTE-U} BS will adjust its duty cycle when one or more Wi-Fi APs turned off, and vice versa. With a large number of Wi-Fi APs, it becomes harder to detect the number accurately using either energy-based or correlation-based approaches. In this work, our goal is to infer the presence of one or more Wi-Fi APs accurately from the collected energy level data using ML algorithms that have been trained on real data. We accomplish this by creating a realistic open lab experimental scenarios using a National Instruments (NI) USRP RIO board with a LTE-U module, five Netgear \mbox{Wi-Fi} APs, and five Wi-Fi clients.

The rest of the paper is organized as follows. Section~\ref{sec:related-work} presents a brief overview of existing studies on ML as applied to wireless networks and LTE/Wi-Fi coexistence in the unlicensed spectrum. Section~\ref{ca} explains the channel access procedure in Wi-Fi using CSMA/CA and the LTE-U duty cycle mechanism. Section~\ref{sm} presents the coexistence system model and the impact of LTE-U and Wi-Fi transmissions on each other. Section~\ref{algo} explains the HD, ED and AC based LTE-U duty cycle adaptation algorithms. Section~\ref{sec:ac-setup} describes the experimental set-up used to measure energy values and gather statistics of the energy level in the presence of one or more \mbox{Wi-Fi} APs. Section~\ref{sec:ML} then evaluates various ML algorithms and chooses the most appropriate one for adjusting the duty cycle based on the collected data. Experimental results are presented in Section~\ref{sec:experimental-results}. Section~\ref{comp} presents the performance (in terms of successful detection, delay and different ML methods) comparison between HD, ED, AC and ML for fixed and different configuration. Finally, Section~\ref{sec:conclusion} concludes the paper with the main contributions and future work in this area. 
\section{Related Work}\label{sec:related-work}
In this section, we briefly discuss (a) the existing work on LTE Wi-Fi coexistence without ML, (b) the use of ML in general wireless networks and (c) the application of ML to LTE Wi-Fi coexistence.
\subsection{Existing work on LTE and Wi-Fi Coexistence}
There has been a significant amount of research, from both academia and industry, on the coexistence of LTE and Wi-Fi that discuss several key challenges such as: Wi-Fi client association, interference management, fair coexistence, resource allocation, carrier sensing, etc. Coexistence scenarios are well studied in simulations for both LAA/Wi-Fi and LTE-U/Wi-Fi deployments \cite{chai2016lte,cano2016unlicensed,chen2016optimizing}.
These papers examine coexistence fairness in varying combinations of detection threshold and duty-cycle. However, the auto-correlation based and energy based methods for spectrum sensing in this coexistence context have not been well studied. Recently, we proposed an energy-based CSAT for duty cycle adaptation in LTE-U ~\cite{sathya2018energy,sathya2018association,vs}, and studied this approach via rigorous theoretical and experimental analyses. The energy-based CSAT algorithm can infer the number of coexisting Wi-Fi APs by detecting the energy level in the channel, which is then used to adjust the duty cycle accordingly. Using a threshold of -42 dBm, the algorithm is able to differentiate between one or two Wi-Fi APs, with a successful detection probability $P_D$ of greater than 80\% and false positive probability $P_{FA}$ of less than 5\%.
Hence, this initial work proved the feasibility of stand-alone energy-based detection, without the need for packet decoding. In our succeeding work, we proposed a novel algorithm that utilizes auto-correlation function (AC) ~\cite{sathya2019auto} to infer the number of active Wi-Fi APs operating in the channel. The AC function is performed on the preamble of a signal to determine if the signal is a Wi-Fi signal. This work further improved the performance of the energy-based approach, with $P_D$ of 0.9 and $P_{FA}$ of less than 0.02, when using an AC threshold $N_E$ of 0.8. In both  \cite{sathya2018energy,sathya2019auto}, the maximum number of Wi-Fi APs considered on the channel was two. In realistic dense deployment scenarios, we can expect more than 2 APs on the same channel. Hence, in this paper we study the performance of ED and AC for more realistic dense deployment scenarios.
\subsection{ML as applied to Wireless Networks}
In \cite{sun2019application}, several state-of-the-art applications of ML
in wireless communication and unresolved
problems have been described. Resource management in the MAC layer, networking and mobility
management in the network layer, and localization in the application layer are some topics that have been identified as being suitable fo ML approaches. Within each of these topics, the  authors provide a survey of the diverse ML based approaches that have been proposed. In  \cite{chen2019artificial,zappone2019wireless}, a comprehensive tutorial has been provided on the use of artificial neural networks-based machine learning for enabling a variety of applications in wireless networks.
In particular, the authors presented an overview of a number of key types of neural networks such as recurrent, spiking, and deep neural networks. For each type, the basic architecture as well as the associated challenges and opportunities have been presented,  followed by an overview of the variety of wireless communication problems that can be addressed using artificial neural networks (ANNs). This work further investigated many emerging applications including unmanned aerial vehicles, wireless virtual reality, mobile edge caching and computing, Internet of Things, and multi-Random Access Technology (RAT) wireless networks. For each application, the author provided the main motivation for using ANNs along with their associated challenges while also providing a detailed example for a use case scenario.
\subsection{ML as applied to LTE Wi-Fi Coexistence}
A learning-based coexistence mechanism for LTE unlicensed based heterogeneous networks (HetNets) was presented in \cite{tan2018learning}. The motivation was to maximize the normalized throughput of the unlicensed band while guaranteeing the Quality of Service (QoS) of users: the authors thus considered the joint resource allocation and network access problem. A two-level framework was developed to decompose the problem into two subproblems, which were then solved using learning-based approaches. The outcome of the proposed solution achieves near-optimal performance and is more efficient and adaptive due to the distributed and learning-based approach. Authors in \cite{bayhantutorial} provide an overview of earning schemes that enable efficient spectrum sharing using a generic cognitive radio setting as well as LTE and Wi-Fi coexistence scenarios. Most LTE-U duty cycle solutions rely on static coexistence parameter configurations, which may not be applicable in real-life scenarios which are dynamic. Hence in \cite{de2019dm}, the author uses the Markov decision process modeling along with a solution based on a ML CSAT algorithm which adapts the LTE duty-cycle ratio to the transmitted data rate, with the aim of maximizing the Wi-Fi and LTE-U aggregated throughput. A ML based approach was proposed in \cite{rastegardoost2018machine}
for a model-free decision-making implementation of opportunistic coexistence of LTE-U with Wi-Fi, which enabled the LTE-U BS to dynamically identify and further exploit white spaces in the Wi-Fi channel, without requiring detailed
knowledge of the Wi-Fi system. By adaptively adjusting the LTE-
U duty cycle to Wi-Fi activity, the proposed algorithm enabled
maximal utilization of idle resources for LTE-U transmissions, while decreasing the latency imposed on Wi-Fi traffic. The proposed approach also provided a means to control the trade-off between LTE-U utilization and Wi-Fi latency in the coexisting networks.

In \cite{maglogiannis2018q}, the author analytically analyzes the LTE-U scheme when it coexists with Wi-Fi and introduces a ML technique that can be used by an LTE-U network to learn the wireless environment and autonomously select the transmission opportunity (TXOP) and muting period configurations that can provide fair coexistence with other co-located technologies. Simulation results show how ML can assist LTE-U in finding optimal configurations and adapt to changes of the wireless environment thus providing the desired fair coexistence. Authors in \cite{maglogiannis2019enhancing} propose a convolutional neural network (CNN) that is trained to
perform identification of LTE and Wi-Fi transmissions which can also identify the hidden terminal effect caused by multiple LTE transmissions, multiple Wi-Fi transmissions, or concurrent LTE and Wi-Fi transmissions. The designed CNN has been trained and validated using commercial off-the-shelf LTE and Wi-Fi hardware equipment. The experimentation results show that the data representation affects the accuracy of CNN. The obtained information from CNN can be exploited by the LTE-U scheme in order to provide fair coexistence between the two wireless technologies.

The above papers on ML in wireless and unlicensed spectrum do not address the problem of accurately identifying the number of Wi-Fi APs which is a crucial first step in addressing fair coexistence for LTE-U/Wi-Fi coexistence. Hence, in this paper, we modify the classical ML approaches to develop algorithms that can identify the number of Wi-Fi APs on air faster and more reliably than existing methods. Our approach is based on collecting data in realistic coexistence environments for both training and testing. We also compare the performance of the ML-based approaches with the more conventional ED and AC methods described above.

\begin{table}[htb!]
\caption{Experimental Set-up Parameters}
\centering
\begin{tabular}{|p{4cm}| p{4cm}|}
\hline\bfseries
\cellcolor{Gray} Parameter&\bfseries \cellcolor{Gray} Value \\ [0.4ex]
\hline
Available Spectrum and Frequency  & 20 MHz and 5.825 GHz \\
\hline
Maximum Tx power for both LTE and \mbox{Wi-Fi} & 23 dBm \\ 
\hline
Wi-Fi sensing protocol & CSMA/CA \\
\hline
Traffic & Full Buffer (Saturation Case) \\
\hline
Wi-Fi \& LTE-U Antenna Type & MIMO \& SISO\\
\hline
LTE-U data and control channel & PDSCH and PDCCH \\
\hline
Type of Wi-Fi Clients & 2 Google Pixel, 1 Samsung, 1 Redmi, and 1 Apple Laptop \\
\hline
\end{tabular}
\label{sim}
\end{table}

\section{Channel Access Procedure for Wi-Fi and LTE-U}\label{ca}
In this section, we discuss the differences in the channel access procedures for Wi-Fi, using CSMA/CA and LTE-U, using the duty cycle mechanism.
\subsection{Wi-Fi CSMA/CA}
\begin{figure}[htb!]
\begin{center}
\includegraphics[width=\linewidth]{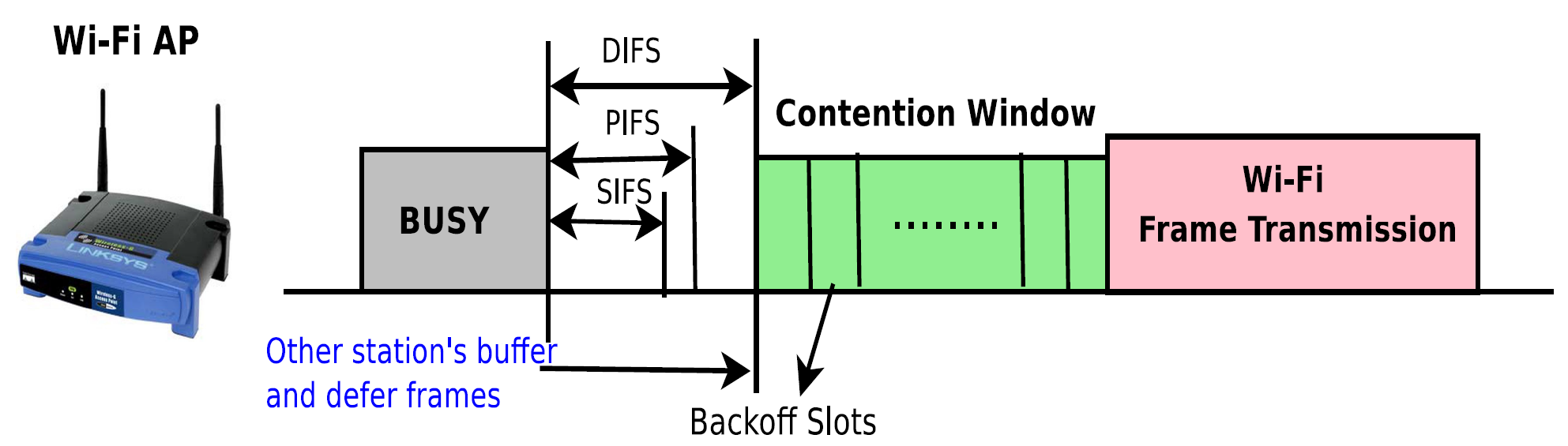}
\caption{Wi-Fi CSMA/CA Transmission}
\label{wifit}
\end{center}
\end{figure}

The Wi-Fi MAC distributed coordination function (DCF) employs CSMA/CA as illustrated in Fig.~\ref{wifit}. Each node attempting transmission must first ensure that the medium has been idle for a duration of DCF Interframe Spacing (DIFS) using the ED and Carrier Sensing (CS) mechanism. If either ED or CS is true, the Clear Channel Assessment (CCA) is set to be busy. If the channel
is idle and the station has not just completed a successful transmission, the station transmits. Otherwise, if the channel is sensed busy during the DIFS sensing period or the station is contending after a successful transmission, the station persists with monitoring the channel until it is measured idle for a DIFS period, then selects a random back-off duration (counted in units of slot time) and counts down. Specifically, a station selects a back-off counter uniformly at random in the range of [0; 2$^i$ $W_0$ - 1] where the value of i (the back-off stage) is initialized to 0 and $W_0$ is the minimum contention window chosen initially. Each failed transmission due to packet collision
results in incrementing the back-off stage by 1 (binary exponential back-off or BEB) and the node counts down from the selected back-off value; \emph{i.e.,} the node decrements the counter every $\sigma$($\mu$s) corresponding to a back-off slot as long as no other transmissions are detected. If during the countdown a transmission is detected, the counting is paused (freeze the back-off counter), and nodes continue to monitor the busy channel until it goes idle; thereafter the medium must remain idle for a further DIFS period before the back-off countdown is resumed for accessing the channel. Once the counter hits zero, the node transmits a packet. When a transmission has been completed successfully, the value of i is reset to 0. The maximum value of back-off stage i is m with the maximum contention window size of $W_m$ and it stays in m-th stage for one more unsuccessful transmission with the same contention window size $W_m$, i.e. the retry limit is 1. The value of $W_0$ and m is determined in the standard. If the last transmission was unsuccessful, the node drops the packet and resets the backoff stage to i = 0. If a unicast transmission is successful the intended receiver will transmit an Acknowledgment frame (ACK) after a Short Interframe Spacing (SIFS) duration post successful reception; the ACK frame structure which consists of preamble and MAC header. The ACK frame chooses the highest basic data rate (6 Mbps, 12 Mbps, or 24 Mbps) for transmitting the MAC header which is smaller than the data rate used for data transmission.

\subsection{LTE-U Duty Cycle}
\begin{figure}[htb!]
\begin{center}
\includegraphics[height=1.8cm,width=9cm]{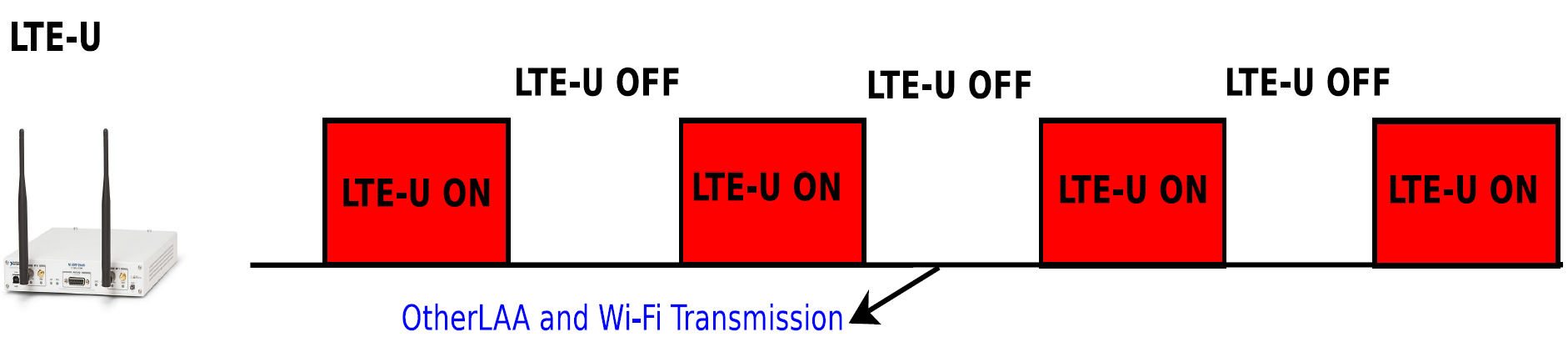}
\caption{LTE-U Duty Cycle Transmission}
\label{lteu12}
\end{center}
\end{figure}

LTE-U uses a duty-cycling approach (i.e. alternating the ON and OFF period, where the LTE BS is allowed to transmit only during the ON duration) where the duty cycle (ratio of ON duration to one cycle period) is determined by perceived Wi-Fi usage at the LTE-U BS, using carrier sensing. During the ON period, the LTE-U BS schedules DL transmissions to UEs, unlike Wi-Fi in which transmissions are governed by the CSMA/CA process. Fig.~\ref{lteu12} shows the LTE-U transmission for the duty cycle of 0.5. 
LTE-U uses the basic LTE subframe structure, \emph{i.e.,} the subframe length of 1 ms; each sub-frame consists of two 0.5 ms slots. Each subframe consists of 14 OFDM symbols of which 1 to 3 are Physical Downlink Control Channel (PDCCH) symbols and the rest are Physical Downlink Shared Channel (PDSCH) data. LTE-U BSs start downlink transmissions synchronized with slot boundaries, for (at least) one subframe (2 LTE slots) duration. After transmission, the intended receiver (or receivers) transmits the ACK on the uplink via the licensed band if the decoding is
successful.

In LTE, a Resource Block (RB) is the smallest unit of radio resource which can be allocated to a user equipment (UE), equal to 180 kHz bandwidth over a Transmission Time Interval (TTI) of one subframe (1 ms). Each RB of 180 kHz bandwidth contains 12 subcarriers, each with 14 OFDM symbols, equaling 168 Resource Elements (REs). Depending upon the modulation and coding schemes (QPSK, 16-QAM, 64-QAM), each symbol or resource element in the RB carries 2, 4 or 6 bits per symbol, respectively. In the LTE system with 20 MHz bandwidth, there are 100 RBs available.

\section{System Model and impact of LTE-U and Wi-Fi on each other}\label{sm}
In this section, we describe the coexistence system model assumed in the paper followed by the mutual impact of LTE-U and Wi-Fi on each other.
\subsection{Coexistence System Model}
We assume a deployment where LTE-U and Wi-Fi are operating on the same unlicensed 20 MHz channel in the 5 GHz band. The LTE-U BS transmits only downlink packets on the unlicensed spectrum, while all uplink transmissions are on the licensed spectrum. Control and data packets are transmitted using PDCCH and PDSCH respectively. The LTE-U BS operates at maximum transmit power using all possible resource blocks and the highest modulation coding scheme (\textit{i.e.}, 64-QAM). We assume that the Wi-Fi APs also operate at maximum transmission power, transmitting a full buffer video traffic. CSMA/CA and duty-cycle adaptation mechanism are used for channel access for Wi-Fi and LTE-U, respectively. Both Wi-Fi and LTE-U follow their respective retransmission schemes such that when a packet transmission is unsuccessful (packet or acknowledgement lost), the packet will be re-transmitted. Finally, we assume that the Wi-Fi APs support both active and passive scanning mode, \emph{i.e.,} both beacon and probe response packets are transmitted by the AP during the association process.
\begin{figure}[t!]
\begin{center}
\includegraphics[width=\linewidth]{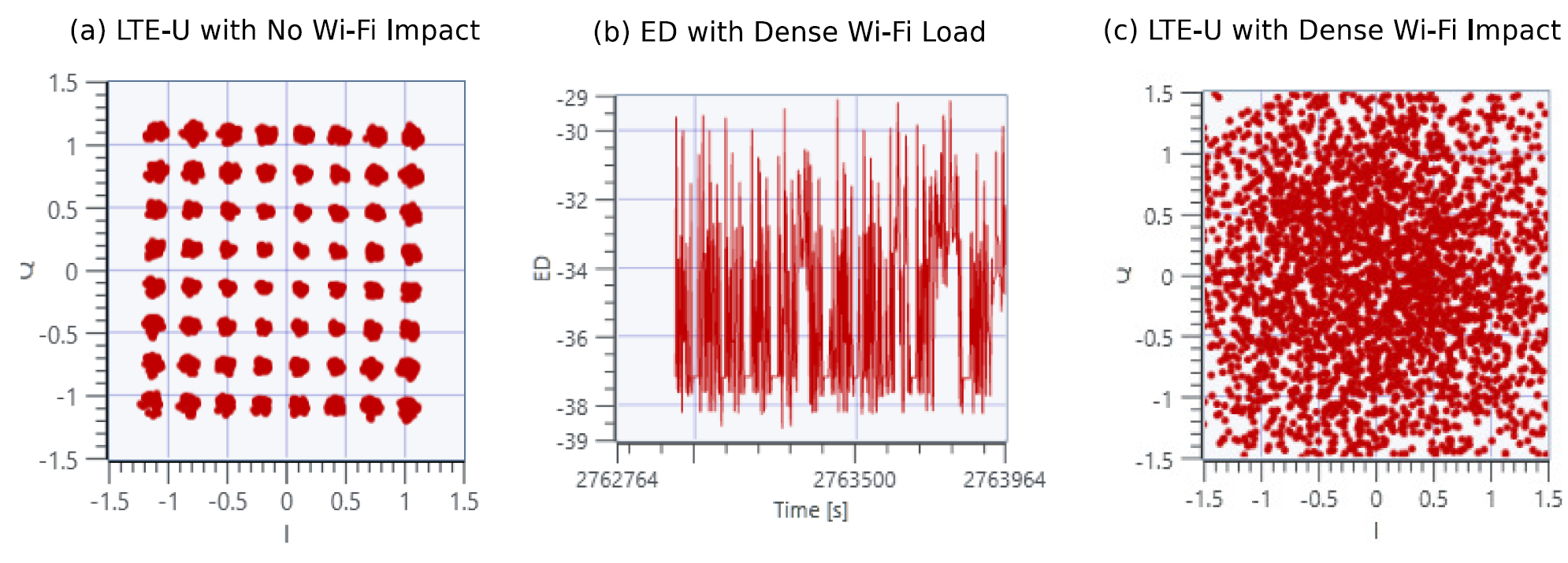}
\caption{Wi-Fi Impact on LTE-U ON Transmission.}
\label{lteu1}
\end{center}
\end{figure}
\begin{algorithm}[H]
\caption{: Header-decoding based LTE-U Scale Back}\label{alg:header}
\textbf{Initialization:}  
$(i)$ $Beacon_i$ = 0 \\
$(ii)$ $Count.detect_{i}$ = 0,  $Count.falsealarm_{i}$ = 0 \\
$(iii)$ $LastTime = 0$, $TimeSlot = 0.512 s$, $Threshold = 4$
\noindent\rule{8.7cm}{0.4pt}
 \While {true} {
 /* A Wi-Fi beacon with BSSID $i$ is detected at time $CurrentTime$ */ \\
 $Beacon_i$ ++; \\
 \If {$CurrentTime - LastTime \geq TimeSlot$} {
    $NumberOfAp = 0$; \\
    \For {$i$ in $Beacon$} {
        \If {$Beacon_i \geq Threshold$} {
            $NumberOfAp$ ++;
        }
        $Beacon_i = 0$; \\
    }
    $LastTime = CurrentTime$; \\
    
    \For {$i$ = 1 to 5} {
        \If  {$i$ Wi-Fi is ON} {
            \If {i == NumberOfAp} {
                $Count.detect_{i}$ ++;
            }
            \Else {
                $Count.falsealarm_{i}$ ++;
            }
        }
    }
 }}
\end{algorithm}
\subsection{Impact of Wi-Fi on LTE-U during the ON period}\label{s1}
In order to observe the impact of Wi-Fi on LTE-U during the ON period (\emph{i.e.,} LTE-U is ON without appropriate sensing of a  Wi-Fi transmission), we deploy a NI based LTE-U BS (Section~V describes the experiment set-up in detail) on channel 165 which is a 20 MHz channel and five Wi-Fi APs on the same channel. Each client is associated with one Wi-Fi AP with full buffer video transmission. Fig.~\ref{lteu1} (a) shows the constellation of received signals when there is no Wi-Fi AP on the channel, that is, LTE-U BS can transmit the data with high modulation coding scheme of 64-QAM. Similarly, Fig.~\ref{lteu1} (b) shows the energy value observed when there are 5 Wi-Fi APs on the same channel, where X-axis represents time and Y-axis represents energy values. Fig.~\ref{lteu1} (c) shows the effect of Wi-Fi transmissions on LTE-U during the ON period, when Wi-Fi APs are unaware of the sudden LTE-U ON cycle starting in the middle of an ongoing Wi-Fi transmission: clearly the constellation is distorted. This clearly points to the inefficient use of the spectrum and the need for the LTE-U BS to sense or learn the medium to identify the number of Wi-Fi APs on the air and scale back its duty cycle accordingly. 
\begin{figure}[t!]
\begin{center}
\includegraphics[width=\linewidth]{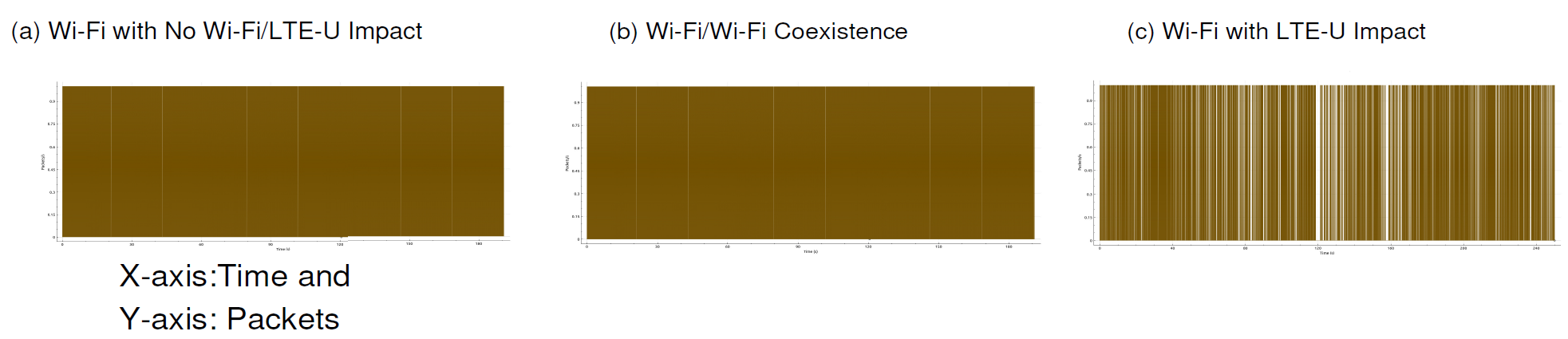}
\caption{LTE-U Impact on Wi-Fi Transmission.}
\label{wifi12}
\end{center}
\end{figure}
\subsection{Impact of LTE-U ON transmission on Wi-Fi Data}\label{s2}
\begin{figure}[t!]
\begin{center}
\includegraphics[width=\linewidth]{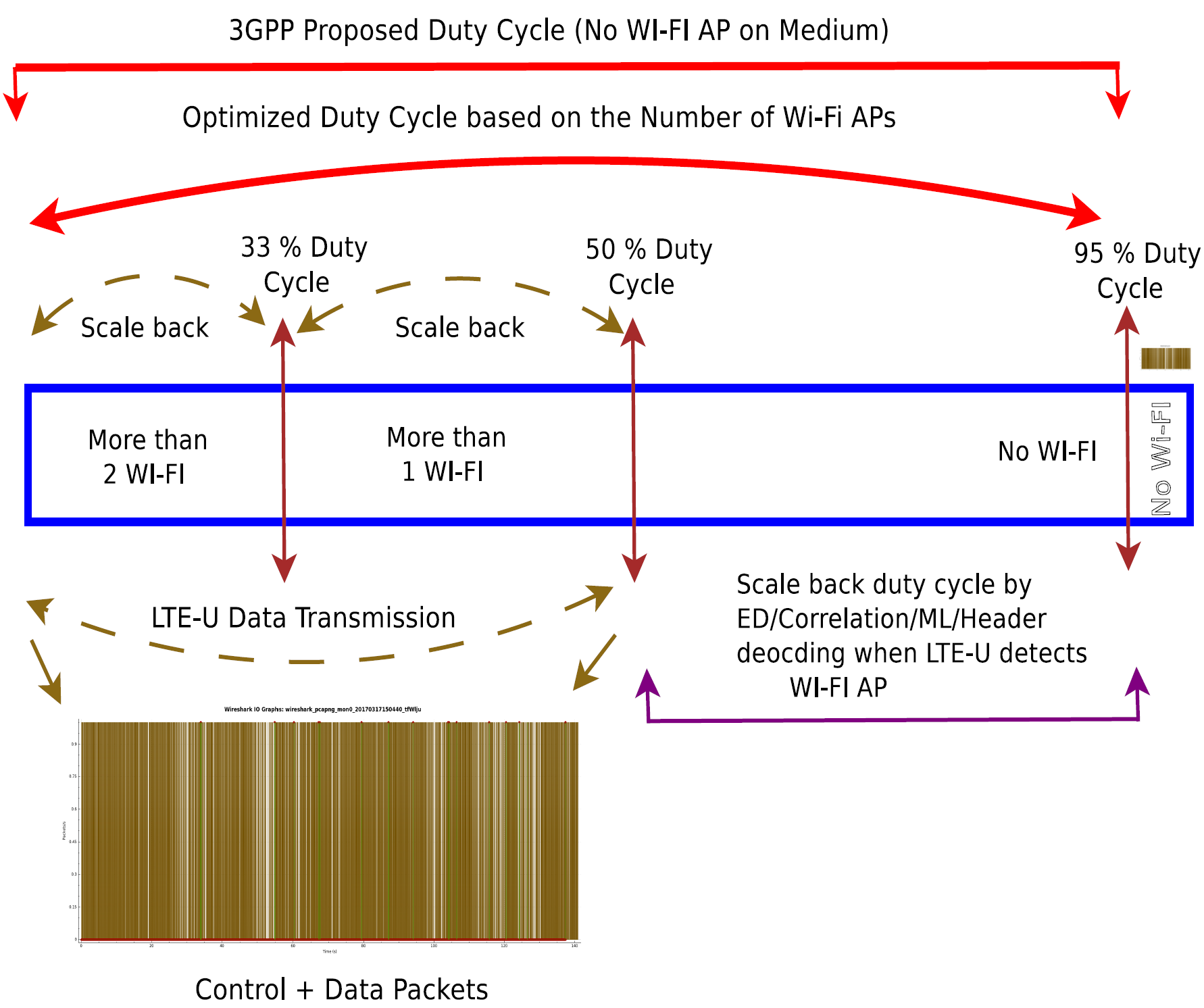}
\caption{LTE-U Duty Cycle Mechanism.}
\label{lteu}
\end{center}
\end{figure}

In case of Wi-Fi/Wi-Fi coexistence where 5 Wi-Fi APs are deployed at the distance of 6F, we observe successful transmission of packets as shown in Fig.~\ref{wifi12} (a) and (b). We see that the CSMA mechanism works well for Wi-Fi/Wi-Fi coexistence, since the number of packets in error with no LTE-U is similar that when Wi-Fi coexists with Wi-Fi. Fig~\ref{wifi12} (c) shows the packet transmission errors when Wi-Fi coexists with a fixed, LTE-U duty cycle: the number of Wi-Fi packets in error increase. To solve the above problem, the LTE-U forum proposed the dynamic CSAT approach \cite{forum,sathya2018energy,sathya2019auto} based on the number of Wi-Fi APs on the same channel. Fig.~\ref{lteu} shows the LTE-U duty cycle adaptation process when detecting a varying number of Wi-Fi APs. When no AP is detected on the channel, an LTE-U BS will operate at the maximum 95\% duty cycle~\cite{forum} (\emph{i.e.,} minimum of 1 ms OFF duration). When one AP is detected (assumed using a predetermined sensing technique), the BS will scale back to 50\% duty cycle (\textit{i.e.}, 20 ms ON time and 20 ms OFF time). If a new Wi-Fi AP starts transmitting, it will contend with the existing AP only during the OFF time which is 50\% of the available medium. Since this is unfair to the Wi-Fi APs, the LTE-U specification recommends scaling the duty cycle back to 33\% when more than one Wi-Fi AP is using the channel. However, there is no specific mechanism proposed to detect the number of coexisting Wi-Fi APs in both sparse and dense deployment scenarios.

\section{Experimental Setup for Machine Learning Based Detection}\label{sec:ac-setup}
Our experimental set-up consists of one LTE-U BS and a maximum of five Wi-Fi APs. To emulate the LTE-U BS, we use the National Instruments USRP 2953-R software defined radio (SDR) which is equipped with the LTE-U radio framework. There are five Netgear Wi-Fi APs and five Wi-Fi clients deployed in a static configuration. The Wi-Fi clients are combination of laptops and smartphones capable of Wi-Fi 802.11 ac connection. As soon as the client connects to the Wi-Fi AP, it starts a live video streaming application to simulate a full-buffer transmission. The experimental setup is shown in Fig.~\ref{exp} and the complete experimental parameters are described in Table~\ref{sim}.

We set the BS and APs to be active in the same 20 MHz channel in the 5 GHz band (\textit{i.e.}, Wi-Fi channel 165 and LTE band 46 EARFCN 53540).
We separated the APs and BS into six cells, with five cells (Cell A, C, D, E, and F) as Wi-Fi cells and one cell (Cell B) as the LTE-U cell. Each Wi-Fi cell consists of one AP and one client, while the LTE-U BS and UE are contained within the same USRP board.

The BS transmits full buffer data at maximum power by enabling all of its resource blocks with the highest modulation coding scheme (\textit{i.e.}, 64-QAM). It operates at a 50\% duty cycle during the experiment, and listens to the configured unlicensed channel during the OFF period for \textit{RF power} and AC measurement. The \textit{RF power} measurement is configured in the LTE block control module of the NI LTE application framework, which outputs energy value as defined in \ref{s3}. The AC function is also configured in the LTE block control module of the same framework and outputs the AC events as defined in \ref{s4}. The energy values observed from Algorithm 2 are given as input to the ML algorithm (explained in detail in Section VII) to classify the number of Wi-Fi APs on the channel. 
Each Wi-Fi AP transmits full buffer downlink data and beacon frames, with occasional probe responses if it receives probe requests for clients in the vicinity. We also ensure that there is no extra interference in the channel from other Wi-Fi APs.

We measure the energy, AC value and ML (same energy value as input to ML) at the LTE-U BS for the following scenarios:
\begin{itemize}
\item \textbf{Scenario 0:} No \mbox{Wi-Fi} APs are deployed and only one LTE-U cell (\textit{i.e.}, Cell B) is deployed.
\item \textbf{Scenario 1:} One \mbox{Wi-Fi} AP (\emph{i.e.,} Cell A) and one LTE-U (\textit{i.e.}, Cell B) is deployed. 
\item \textbf{Scenario 2:} Two \mbox{Wi-Fi} APs (\textit{i.e.}, Cell A \& C) and one LTE-U (\textit{i.e.}, Cell B) are deployed.
\item \textbf{Scenario 3:} Three \mbox{Wi-Fi} APs (\textit{i.e.}, Cell D, E, \& F) and one LTE-U (\textit{i.e.}, Cell B) are deployed.
\item \textbf{Scenario 4:} Four \mbox{Wi-Fi} APs (\textit{i.e.}, Scenario 1: Cell A, Scenario 3: Cell D, E, \& F) and one LTE-U (\textit{i.e.}, Cell B) are deployed.
\item \textbf{Scenario 5:} Five \mbox{Wi-Fi} APs (\textit{i.e.}, Cell A, C, D, E, \& F) and LTE-U (\textit{i.e.}, Cell B) are deployed.
\end{itemize}

In all scenarios, Cell B measures the energy and AC values during the LTE-U OFF period, while the rest of the Wi-Fi cells are transmitting full buffer downlink transmission. We also vary the distances and the LOS and NLOS environment of each cell. In NLOS setup, the wall act as a obstruction between the LTE-U and Wi-Fi APs. We measure the received Wi-Fi AP signals at the LTE-U BS for different 6 feet (For example in Scenario 5, where all the 5 Wi-Fi APs placed at 6 feet from the LTE-U BS), 10 feet and 15 feet distances. Our previous work focused only on detecting Scenarios 1 and 2 (\textit{i.e.}, 1 and 2 Wi-Fi APs coexisting with LTE-U)~\cite{sathya2018energy, sathya2019auto}. Also, we demonstrated that Scenario 0 can be easily distinguished from other scenarios~\cite{vs}.

\section{LTE-U Duty Cycle Adaptation Algorithms}\label{algo}
\begin{figure}[htb!]
\begin{center}
\includegraphics[height=7.6
cm,width=9cm]{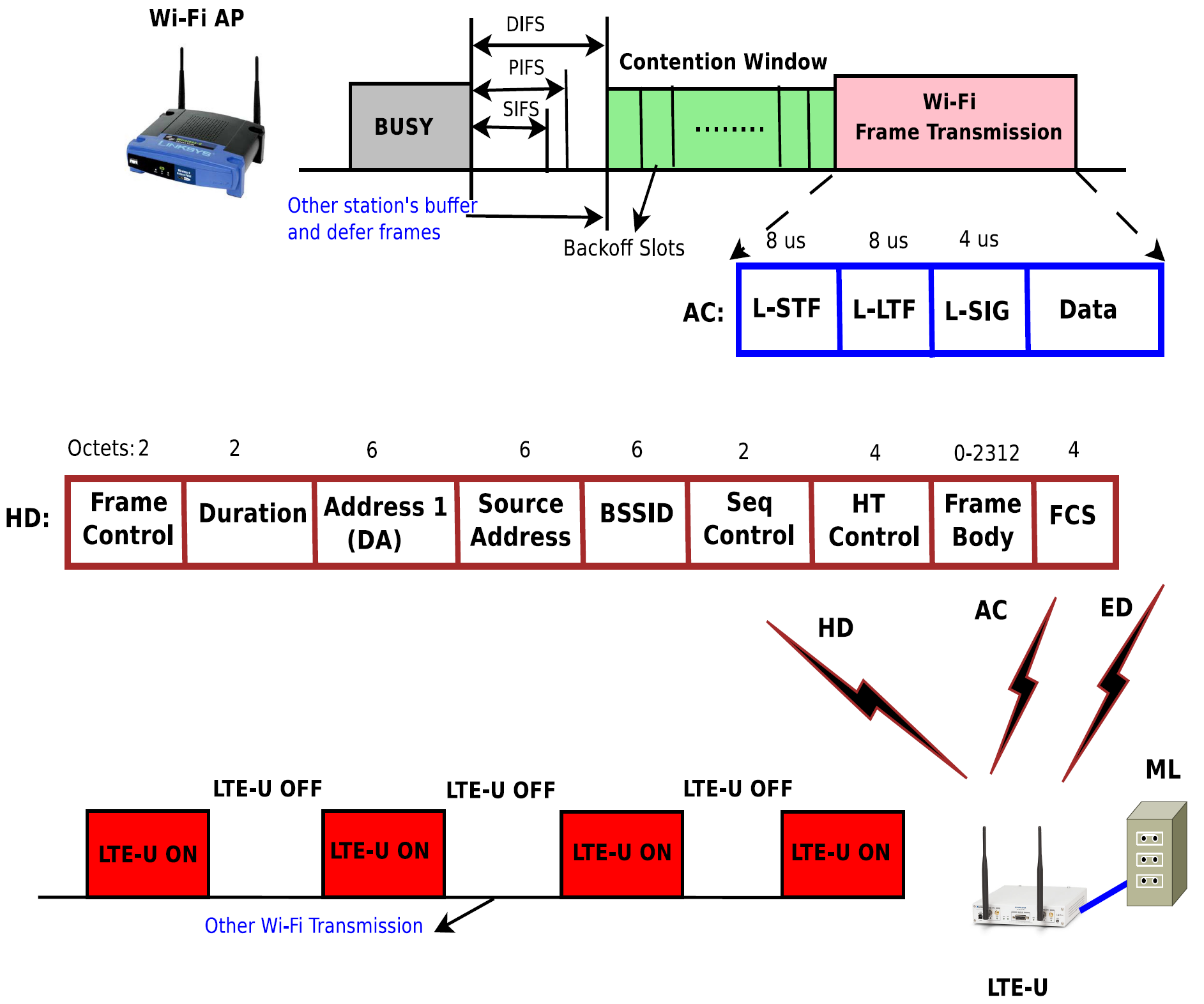}
\caption{LTE-U Duty Cycle Adaptation Algorithm.}
\label{dc1}
\end{center}
\end{figure}

In order to solve the problems identified in the previous section, we propose header (HD), energy (ED) and auto-correlation (AC) based detection algorithms for a dense deployment scenario to identify the number of Wi-Fi APs on the channel. Fig.~\ref{dc1} explains how different sensing algorithms work based on the known Wi-Fi packet structure. 

\subsection{Header-Decoding based LTE-U duty cycle adaptation algorithm}
We assume that there is either a common preamble \cite{Quantenna,att} between the LTE-U and Wi-Fi systems or the LTE-U BS has a full Wi-Fi decoder that will allow it to decode the Wi-Fi MAC header and hence obtain the BSSID. Doing so, one can accurately detect the number of Wi-Fi APs on the channel and hence header-based decoding is the most accurate method compared to energy, auto-correlation, and ML. However, the decision algorithm to adapt the duty cycle needs to be designed carefully to avoid misclassification.
We define a simple algorithm shown in Algorithm~\ref{alg:header},  to classify the number of active Wi-Fi APs at each time slot.
In brief, the algorithm counts the number of beacon of each uniquely identifiable BSSID, for a defined time slot. Since we can expect that an AP in a real deployment may hop between channels frequently, it is important to collect beacons for a longer period of time rather than deciding based on just one beacon. We initially set a time slot of 10 beacons
(\textit{i.e.}, 1.024 s) and count the number of beacons for each BSSID in the time slot. We set a threshold of 9 beacons for an AP to be considered as active, this means that there is 90\% confidence that the AP is actually active. 
The length of the time slot determines the inference delay, hence one would like this delay to be as small as possible. We reduced the time slot to 5 beacons (0.512 s), but to get the same accuracy we need to set the threshold to 4 beacons which means that the confidence rate is at a lower 80\%. Thus, with a slightly lower confidence rate, we can reduce the inference time to half without compromising the detection accuracy.

\begin{algorithm}[H]
\caption{Energy Based LTE-U Scale Back}\label{alg:energy}
\textbf{Input:} $\alpha_1, \alpha_2, \alpha_3, \alpha_4, \alpha_5$ \\
\textbf{Initialization:} $(i)$ $\alpha_6 = \infty$ \\
\hspace{2.5cm} $(ii) Count.detect_i$ = 0,  $Count.falsealarm_i$ = 0 \\
\noindent\rule{8.7cm}{0.4pt}
\While {true} {   
    /*  Received $Avg(Energy Level)$ over one second  */ \\
        \For {$i$ = 1 to 5} {
            \If {$i$ Wi-Fi is ON} {
                \eIf {$\alpha_i \leq$ Avg(Energy Level) $\leq \alpha_{i+1}$} {
                    $Count.detect_i$ ++;
                }
                {
                    $Count.false.alarm_i$ ++; 
                }
            }
        }
    
}
\end{algorithm}

\subsection{Energy based LTE-U duty cycle adaptation algorithm}\label{s3}
The experiment setup is shown in Fig.~\ref{exp}. We measure the received energy at the LTE-U BS for different distances between the LTE-U BS and Wi-fi APs and obtain histograms of the measured signal when one or more Wi-Fi APs are transmitting at 6, 10 and 15 feet from the LTE-U BS. We then fit the measured histograms to probability distribution functions as described in \cite{sathya2018energy} to develop a classification algorithm.
In Algorithm \ref{alg:energy}, an
energy-based detection listens to the energy level in the channel and according to a set threshold \cite{sathya2018energy}, decides whether to scale back the duty cycle or not. Since the measured energy threshold depends on the the number of detected Wi-Fi APs, the choice of threshold is important to the algorithm.
Finally, we implement the algorithm in the LTE-U BS NI hardware and validate it experimentally.

First, we modify the NI LTE application framework to measure \textit{RF power} during the LTE-U OFF period. The collected energy values are then averaged over one second time duration and used for algorithm input. If the averaged energy value is greater than the specified threshold $\alpha_1$, \textit{i.e.}, if energy value $\geq \alpha_1$ then there is a possibility of Wi-Fi packets (beacon, probe request, probe response, data, or ACK) transmitted in the channel. The BS then can declare whether one, two, three, four, or five AP is present, based on the other thresholds: $\alpha_2$, $\alpha_3$, $\alpha_4$, $\alpha_5$ (\textit{e.g.}, if $\alpha_3 \leq \text{energy value} \leq \alpha_4$ then there are 4 APs in the channel). By keeping count of correct and incorrect decisions made by the algorithm, we calculate the probability of correct detection and false positive on predicting the number of Wi-Fi APs in the unlicensed spectrum. These probability values are used as a metric to determine the performance of the threshold, such that we pick a set of threshold with high probability of correct detection and low probability of false positive.

\begin{algorithm}[H]
\caption{: Auto-correlation Based LTE-U Scale Back}\label{alg:ac}
\textbf{Input:} $th_\rho$, $R$ \\
\textbf{Initialization:}  
$Count.detect_i$ = 0, $Count.falsealarm_i$ = 0 \\
\noindent\rule{8.7cm}{0.4pt}
\While {true} {       
    /*  Received $T$ number of $AC$ values over one second */ \\
    \For {i = 1 to 5} {
        \If  {i Wi-Fi is ON}{
            $Signal = 0$; \\
            \For {t = 1 to T} { 
                \If {$AC_t$ $\geq$ $th_{\rho}$} {
                    $Signal$ ++;
                }
            }
            $ratio$ = $\frac{Signal}{T}$; \\
            
            \eIf { $ratio \leq R_i$ } {
                $Count.detect_i$ ++;
            }
            {
                $Count.falsealarm_{i}$ ++; 
            }
        }
    }
       }
\end{algorithm}

\subsection{AC based LTE-U duty cycle adaptation algorithm}\label{s4}
\begin{figure*}[htb!]
\begin{center}
\includegraphics[totalheight=11cm,width=12.5cm]{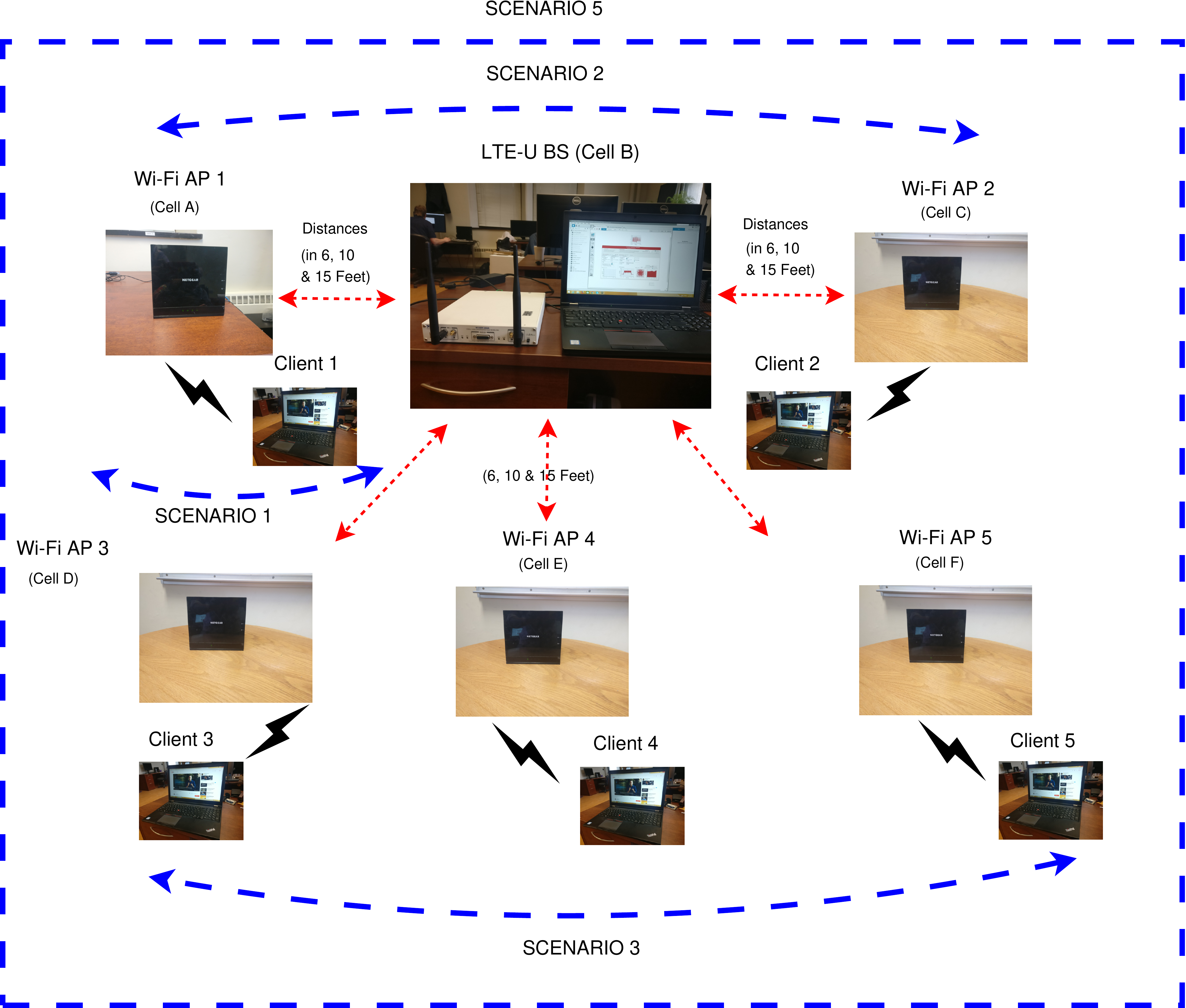}
\caption{LTE \mbox{Wi-Fi} Co-existence Experimental Setup.}
\label{exp}
\end{center}
\end{figure*}

In the same experiment setup as shown in Fig.~\ref{exp}, we count the total number of AC events that are above a threshold for every one second over the
duration of 90 seconds. We measure the total number of events above the AC threshold at the LTE-U BS for 6, 10 and 15 feet distances. Then, we observe the PDF distribution of the number of AC events above the threshold \cite{sathya2019auto} for Scenario 0 to 5 described above. We make use of this key observation to develop a classification algorithm (\emph{i.e.,} Algorithm~\ref{alg:ac}) for both LOS and NLOS scenarios. The algorithm uses AC functions and optimal thresholds to determine the number of Wi-Fi APs in the channel, therefore the selection of threshold is also important and will be shown in this section. We implement the algorithm in the LTE-U BS hardware and validate it experimentally. The AC function is performed at LTE-U BS to sense the spectrum for Wi-Fi preamble signals (\emph{i.e.,} L-STF). The output of the function is an AC value which determine the likelihood that the signal is a Wi-Fi preamble. We observed on many experiments, that the threshold $th_\rho$ of 0.25 is sufficient to determine that the captured signal is a Wi-Fi signal (beacon, probe request, probe response, data, or ACK). Using the threshold, we predicted the number of Wi-Fi signals in every one second period. Next, we calculate the ratio \cite{sathya2019auto} and then compared to $R_i$ which is a threshold determined during a preliminary experiment with $i$ Wi-Fi AP and no LTE-U on the channel. The $R_i$ is determined such that the true positive rate is as high as possible and false positive rate is as low as possible during the preliminary experiment. Since it is not possible for the observed ratio to be higher than $R_i$, we set a correct prediction that $i$ Wi-Fi AP is present in the channel if the ratio is less than or equal to the threshold $R_i$, and false prediction otherwise.

\section{ML Algorithms for LTE-U Duty Cycle Adaption}\label{sec:ML}

ML models enable us to replace heuristics with more robust and general alternatives. For the problem of distinguishing between different numbers of \mbox{Wi-Fi} APs, we train a model to detect a pattern in the signals instead of finding a specific energy threshold in a heuristic manner. The state-of-the-art ML models leverage the unprecedented performance of neural network models that are able to surpass human performance on many tasks, for example, image recognition~\cite{he2015delving}, and help us answer complex queries on videos~\cite{krishnan2018deeplens}. This efficiency is a result of large amounts of data that can be collected and labeled as well as usage of highly parallel hardware such as GPUs or TPUs~\cite{jouppi2017datacenter,chetlur2014cudnn}. In the work described in this paper, we train our neural network models on NVidia GPUs and collect enough data samples that enable our models to achieve high accuracy. Our major task is a classification problem to distinguish between zero, one, two, three, four, or five \mbox{Wi-Fi} BSSs. 

We consider machine learning models that take time-series data of width $w$ as input, giving an example space of $\mathcal{X} \in \mathcal{R}^{w}$, where $\mathcal{R}$ denotes the real numbers. Our discrete label space of $k$ classes is represented as $\mathcal{Y} \in \{0,1\}^k$. For example, $k=3$ classes, enables us to distinguish between 0, 1, and 2 \mbox{Wi-Fi} APs. Machine learning models represent parametrized functions (by a weight vector $\theta$) between the example and label spaces $f(x;\theta): \mathcal{X} \mapsto \mathcal{Y}$. The weight vector $\theta$ is iteratively updated during the training process until the convergence of the training accuracy or training loss (usually determined by very small changes to the values despite further training), and then the final state of $\theta$ is used for testing and real-time inference.

\subsection{Data preparation}
The training and testing data is collected over an extended period of time with a single scenario taking about 8 hours. For ease of exposition, we consider the case with one and two \mbox{Wi-Fi} APs. We collect data for each \mbox{Wi-Fi} AP independently and store the two datasets in separate files. Each file contains more than 2.5 million values and the total raw data size in CSV format is of about 60 MB. Each file is treated as time-series data with a sequence of values that are first divided into chunks. We overlap the time-series chunks arbitrarily by three-fourths of their widths $w$. For example, for chunks width $w=128$, the first chunk starts at index 0, the second chunk is formed starting from index 32, the third chunk starts at index 64, and so on. This is part of our data augmentation and a soft guarantee that much fewer patterns are broken on the boundary of chunks. The width $w$ of the (time-series data) chunk acts as a parameter for our ML model. It denotes the number of samples that have to be provided to the model to perform the classification. The longer the time-series width $w$, the more data samples have to be collected during inference. The result is higher latency of the system, however, the more samples are gathered, the more accurate the predictions of the model. On the other hand, with smaller number of samples per chunk, the time to collect the samples is shorter, the inference is faster but of lower accuracy. We elaborate more on this topic in Section~\ref{sec:experimental-results}.

The collection of chunks are shuffled randomly. We divide the input data into training and test sets, each 50\% of the overall data size. The aforementioned shuffling ensures that we evenly distribute different types of patterns through the training and test sets so that the classification accuracy of both sets is comparable. Each of the training and test sets contain roughly the same number of chunks that represent one or two \mbox{Wi-Fi} APs. We enumerate classes from 0. For the case of 2 classes (either one or two \mbox{Wi-Fi}s), we denote by \textit{0} the class that represents a single \mbox{Wi-Fi} AP and by \textit{1} the class that represents 2 \mbox{Wi-Fi} APs. Next, we compute the mean $\mu$ and standard deviation $\sigma$ only on the training set. We check for outliers and replace the values that are larger than $4\sigma$ with the $\mu$ value (e.g., there are only 4 such values in class \textit{1}). 

The data for the two classes have different ranges (from about -45.46 to -26.93 dBm for class \textit{0}, and from about -52.02 to about -22.28 dBm for class \textit{1}). Thus, we normalize the data $D$ in the standard way: $ND = \frac{(D - \mu)}{\sigma}$, where $ND$ is the normalized data output, $\mu$ and $\sigma$ are the mean and standard deviation computed on the training data. We attach the appropriate label to each chunk of the data. The overall size of the data after the preparation to detect one or two \mbox{Wi-Fi} APs is about 382 MB, where the \mbox{Wi-Fi} APs are on opposite sides of the \mbox{LTE-U} BSS and placed at 6 feet distance from the \mbox{LTE-U} BSS). We collect data for many more scenarios and present them in Section~\ref{sec:experimental-results}. The final size of the collected data is 3.4 GB. 

For training, we do not insert values from different numbers of \mbox{Wi-Fi} APs into a single chunk. The received signal in the \mbox{LTE-U} BSS has higher energy on average for more \mbox{Wi-Fi} APs, thus there are differences in the mean values for each dataset. 
Our data preparation script handles many possible numbers of \mbox{Wi-Fi} APs and generates the data in the format that can be used for model training and inference (we follow the format for datasets from the UCR archive). In the future, we plan on gathering additional data samples for more Wi-Fi APs and making the dataset more challenging for classification.

\subsection{Neural network models: FC, VGG and FCN}
\label{sec:fcn-neural-nets}
Our data is treated as a uni-variate time-series for each chunk. There are many different models proposed for the standard time-series benchmark~\cite{chen2015ucr}. 

First, we test \textit{fully connected (FC)} neural networks. For simple architectures with two linear layers followed by the ReLU non-linearity the maximum accuracy achieved is about 90\%. More linear layers, or using other non-linearities (e.g. sigmoid) and weight decays do not help to increase the accuracy of the model significantly. Thus, next we extract more patterns from the data using the convolutional layers. 

Second, we adapt the \textit{VGG} network~\cite{simonyan2014very} to the one dimensional classification task. We changed the number of weight layers to 6 (we also tested 7, 5, and 4 layers, but found that 6 gives the highest test accuracy of about 99.52\%). However, the drawback is that with fewer convolutional layers, the fully connected layers at the end of \textit{VGG} net become bigger to the point that it hurts the performance (for 4 weight layers it drops to about 95.75\%). This architecture gives us higher accuracy but is rather difficult to adjust to small data.\footnote{The dimensionality of the data is reduced slowly because of the small filter of size 3.}

Finally, one of the strongest and flexible models called \textit{FCN} is based on convolutional neural networks that find general patterns in time-series sequences~\cite{wang2017time}. The advantages of the model are: simplicity (no data-specific hyper-parameters), no additional data pre-processing required, no feature crafting required, and significant academic and industrial effort into improving the accuracy of convolutional neural networks~\cite{dziedzic2019band, lavin2016fast}. 

The architecture of the FCN network contains three blocks, where each of them consists of a convolutional layer, followed by batch normalization $f(x) = \frac{x - \mu}{\sqrt{\sigma^2 + \epsilon}}$ (where $\epsilon$ is a small constant added for numerical stability) and ReLU activation function $y(x) = \max(0, x)$. There are 128, 256, and 128 filter banks in each of the consecutive 3 layer blocks, where the sizes of the filters are: 8, 5, and 3, respectively. We follow the standard convention for Convolutional Neural Networks (CNNs) and refer to the discrete cross-correlation operation as convolution. The input $x$ to the first convolution is the time-series data chunk with a single channel $c$. After its convolution with $f$ filters, with filters denoted as $y$, the output feature map $o$ has $f$ channels. For training, we insert $s=32$ time-series data chunks into a mini-batch. We have $j \in f$ and the discrete convolution~\cite{vasilache2014fast} that can be expressed as:
\begin{align}
    o &= x * y
\end{align}
and in the Einstein notation:
\begin{align}
    o_{(s,j)} &= \sum_{i \in c} x_{(s,i)} \cdot y_{(j,i)} 
\end{align}

\subsection{ML models from scikit-learn}
\label{sec:sklearn-ml}
To diversify the machine learning models used in our comparison, we select the most popular models from the scikit-learn (also denoted as \textit{sklearn}) library~\footnote{https://scikit-learn.org/stable/index.html}. The library exposes classical machine learning algorithms implemented in Python. This is a common tool used for science and engineering. We run our experiments using \textit{sklearn} version 0.19.1 with Python 3.6. We analyze how the following models perform on our WiFi data and report their test accuracy. The decision tree is a simple classifier that learns decision rules inferred from the data features. The deeper the tree, the more complex the decision rules and the fitter the model. The decision tree classifier achieves accuracy of 79.46\% for the task of distinguishing between one or two \mbox{Wi-Fi} APs. The AdaBoost~\cite{multiClassAdaBoost} classifier is one of the best out-of-the-box models in the \textit{sklearn} library that creates an ensemble of classifiers. In our experiments, AdaBoost begins by fitting a decision tree classifier on the original dataset and then fits additional decision tree classifiers on the same dataset but where the weights of incorrectly classified instances are modified such that subsequent classifiers focus more on difficult cases. It is tuned by adjusting the maximum number of the decision tree classifiers used. AdaBoost achieves accuracy  of 94.57\%. Random Forest is an averaging algorithm based on randomized decision trees. Its test accuracy is 79.87\%. 
We find that the best tested model from the \textit{sklearn} library is AdaBoost. The highest test accuracy achieved for AdaBoost for the standard case with two \mbox{Wi-Fi} APs is worse by about 5\% when compared to the overall best FCN model (described in section~\ref{sec:fcn-neural-nets}), which achieves accuracy of 99.38\% for the same configuration (with 2 \mbox{Wi-Fi} APs, 512 chunk size, NLOS, and 6 feet distance). For more than 5 classes, Random Forest model achieves higher accuracy than AdaBoost.

\subsection{Time-series specific models}
\label{sec:boss-vs}
BOSS in Vector Space (BOSS VS) model~\cite{boss-vs} is a time-series classification algorithm, whose properties make it suitable for our task. This algorithm is characterized by fast inference, tolerance to noise that enable us to achieve high test accuracy, moderate training time, which allows for periodic model updates. Moreover, BOSS VS achieves best test accuracy for repetitive and long time-series data. Within the time-series specific models, we also compared to WEASEL~\cite{WEASEL}, which yielded lower test accuracy despite much longer training time.

We run the BOSS VS time-series specific model for the NLOS 6 feet case. Other time-series models train much longer (in the order of days) on our large (a few GBs) time-series data or do not fit even into 128 GB of RAM memory provided. We observe that from 2 to up to 4 WiFi APs, the performance of the BOSS VS model is on-par with the performance of FCN model. However, for the scenario where we have to distinguish between 0 to 5 WiFi APs, the accuracy of the FCN model is higher by about 7\%. One concern with the BOSS VS model is that we have to use a machine with 128 GB of RAM to train the model and for larger data sizes, the out of memory exception is thrown as well (the model is implemented in Java). For the FCN, we are able to scale to arbitrary amount of data. Based on the thorough experimental analysis, we see the FCN model and other neural network based models as the most accurate and scalable models that can be used to predict the number of Wi-Fi APs.

\subsection{FFT compression}

We use the FFT-based convolution with compression proposed in~\cite{dziedzic2019band} and here describe its essential component. We express input $x$ and filter $y$ as discrete functions that map tensor index positions $n$ to values $x[n]$. Their corresponding Fourier  representation re-indexes tensors in the spectral domain:
\[
F_x[\omega] = F(x[n]) ~~~~~~~  F_y[\omega] = F(y[n])
\] 
This mapping is invertible $x = F^{-1}(F(x))$. Convolutions in the spectral domain correspond to element-wise multiplications:
\[
x * y = F^{-1}(F_x[\omega] \cdot F_y[\omega])
\]
For natural data, such as time-series data, a substantial portion of the high-frequency domain is close to 0. This observation allows us to compress the data.

Let $M_c[\omega]$ be a discrete indicator function defined as:
\[
M_c[\omega] = \begin{cases}
1, \omega \le c\\
0,\omega > c
\end{cases}
\]
$M_c[\omega]$ is a mask that limits the input data and filters to a certain \emph{band} of frequencies.
The FFT-based convolution with compression is defined as follows:
\begin{align}
 x *_c y & = F^{-1}\{(F_x[\omega] \cdot M_c[\omega]) \cdot (F_y[\omega] \cdot M_c[\omega])\}
 \label{eq:fft-based-conv}
\end{align}

 The mask $M_c[\omega]$ is applied to both the signal $F_x[\omega]$ and filter $F_y[\omega]$ (in equation~\ref{eq:fft-based-conv}) to indicate the compression of both arguments. 

\section{Experimental results}\label{sec:experimental-results}
In this section we discuss the model training, inference and transition between different classes. The code for our project can be found on github:~\url{http://bit.ly/2Ob5kAr}.

\subsection{Training and Inference}
Each model is trained for at least 100 epochs. We experiment with different gradient descent optimization algorithms, e.g. Stochastic Gradient Descent (SGD) and Adaptive Moment Estimation (Adam)~\footnote{A very good explanation can be found here: http://bit.ly/2Y9XaQ8}. For the SGD algorithm, we grid search for the best initial learning rate and primarily use 0.0001. The learning rate is reduced on plateau by 2X after 50 consecutive iterations (scheduled patience). SGD is used with momentum value 0.9. We use standard parameters for the Adam optimization algorithm. The batch size is set to $s=32$ to provide high statistical efficiency. The weight decay is set to 0.0001. For our neural network models, the dataset is relatively simple. The \mbox{Wi-Fi} data can be compared in its size and complexity to the MNIST dataset~\cite{deng2012mnist} or to the GunPoint series from the UCR archive~\cite{chen2015ucr}.

\subsection{Time-series width}
\begin{figure}[htb!]
\begin{center}
\includegraphics[width=8cm, height = 5.2cm]{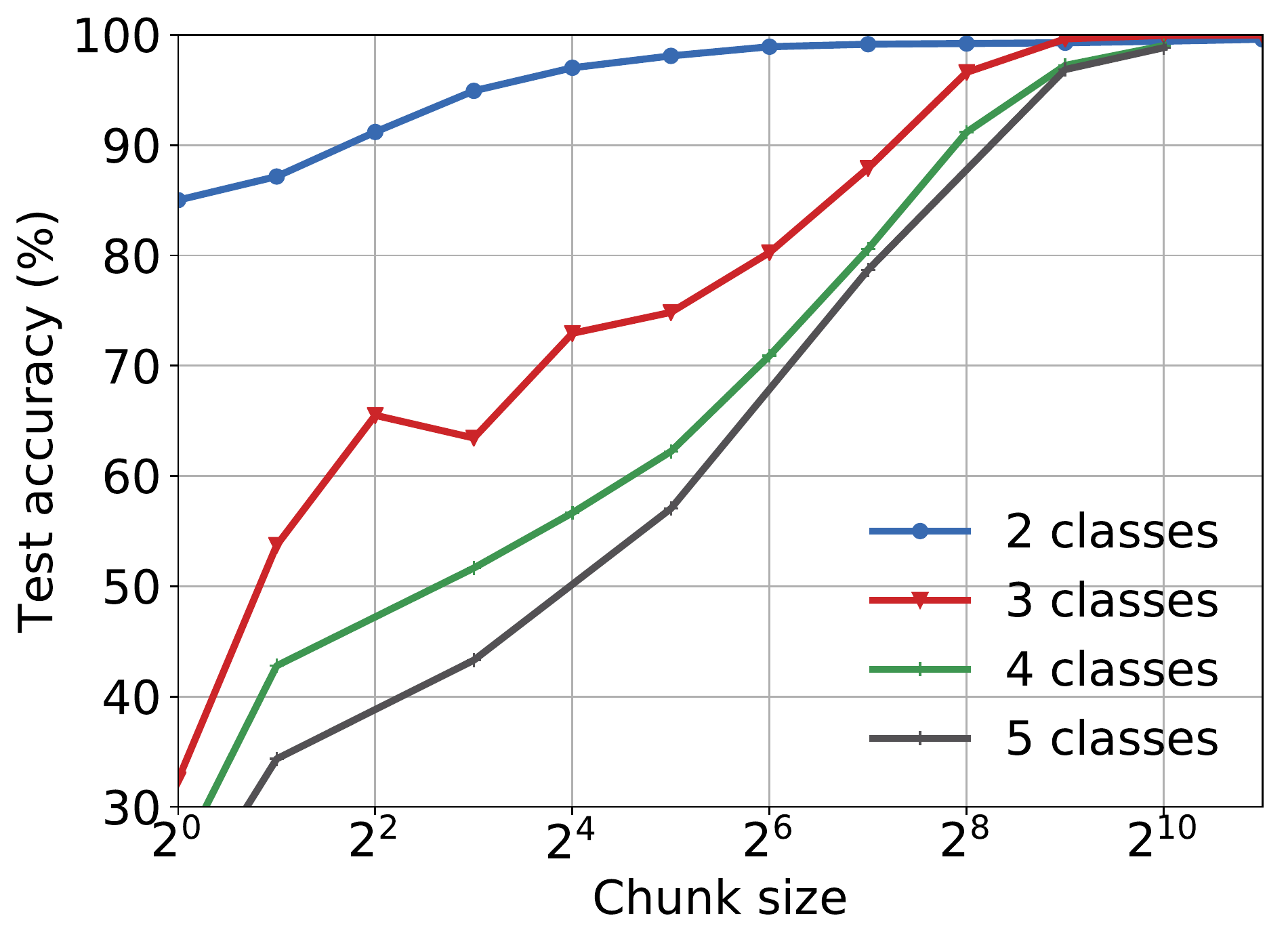}
\caption{The test accuracy (\%) for a model trained and tested for a given chunk size (ranging from 1 to 2048) to distinguish between 2 classes (either 1 or 2 \mbox{Wi-Fi} APs), 3 classes (distinguish between 0, 1, or 2 \mbox{Wi-Fi} APs), 4 classes (distinguish between 0, 1, 2, or 3 \mbox{Wi-Fi} APs), and 5 classes (distinguish between 0, 1, 2, 3 or 4 \mbox{Wi-Fi} APs) }
\label{fig:test-accuracy-chunk-size}
\end{center}
\end{figure}

The number of samples collected per second by the LTE-U BS is about 192. The inference of a neural network is executed in milliseconds and can be further optimized by compressing the network. The final width of the time-series chunk imposes a major bottleneck in terms of the system latency. The smaller the time-series chunk width $w$, the lower the latency of the system. However, the neural network has to remain highly accurate despite the small amount of data provided for its inference. Thus, we train many models and systematically vary the chunk width $w$ from 1 to 2048 (see Fig.~\ref{fig:test-accuracy-chunk-size}). In this case, each model is trained only for the single scenario (placement of the \mbox{Wi-Fi} APs) and with zero, one, two, or three active \mbox{Wi-Fi} APs. When we decrease the chunk sizes to the smaller chunk consisting of a single sample, the test accuracy deteriorates steadily down to the random choice out of the 3 or 4 classes (accuracy of about 33\% and 25\%, respectively) and for the 2 classes, its performance is very close to the ED (Energy-based Detection) method.

\begin{figure}[htb!]
\begin{center}
\includegraphics[width=1.0\linewidth]{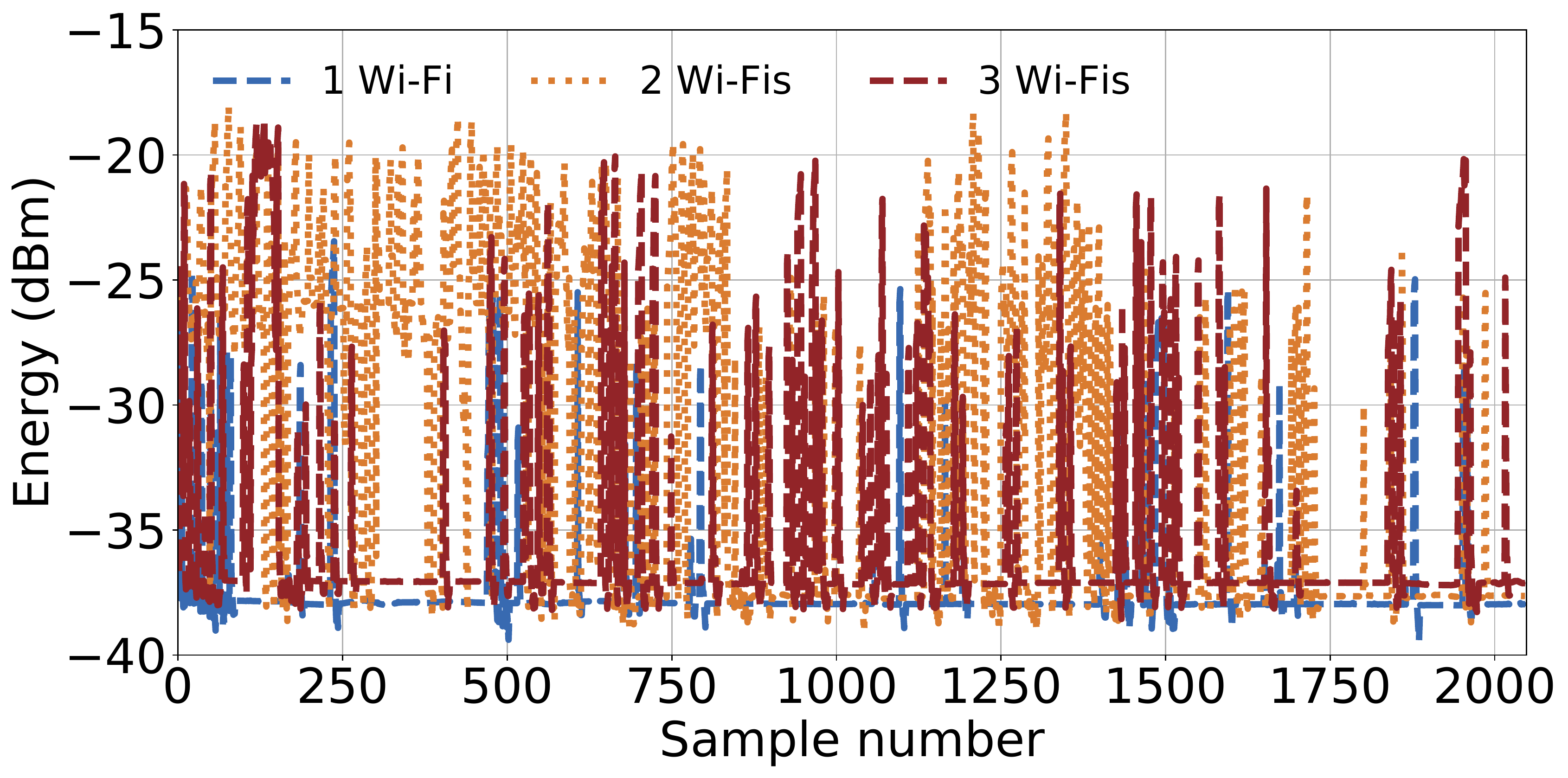}
\caption{\textbf{Number of \mbox{Wi-Fi} APs.} The values of the energy (in dBm) captured for 2048 samples in \mbox{LTE-U} BS while there are 1 Wi-Fi, 2, and 3 \mbox{Wi-Fi}s scenarios at 6 Feet, NLOS. The more \mbox{Wi-Fi} APs active, the more energy picks we observe.}
\label{fig:energy_values_number_of_wifis}
\end{center}
\end{figure}

\begin{figure}[htb!]
\begin{center}
\includegraphics[width=1.0\linewidth]{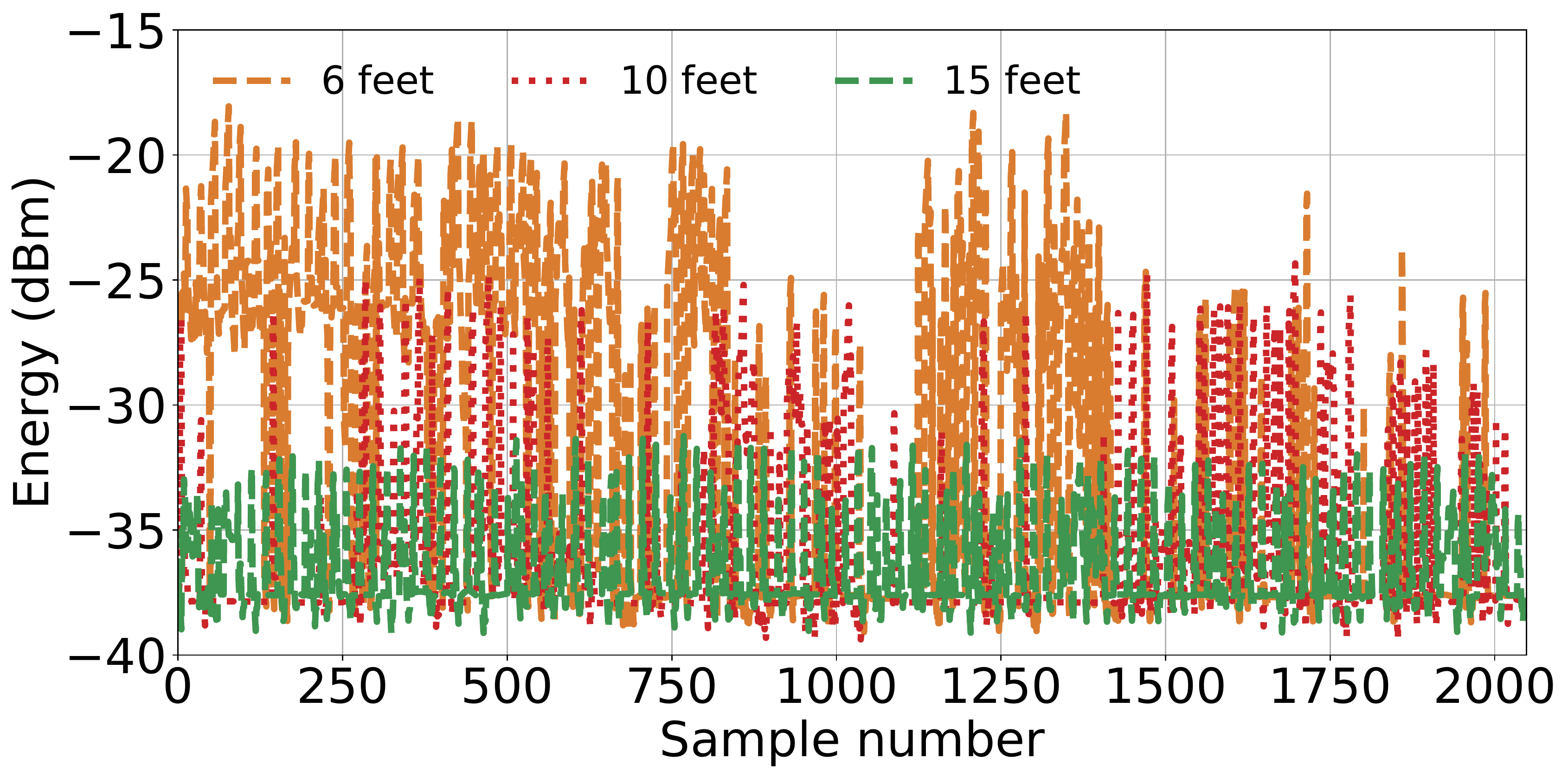}
\caption{\textbf{Distances from LTE-U.} The values of the energy (in dBm) captured for 2048 samples in \mbox{LTE-U} BS while there are 2 \mbox{Wi-Fi} APs at 6, 10, and 15 Feet, NLOS. The closer the \mbox{Wi-Fi} APs are to the LTE-U, the higher energy is captured.}
\label{fig:energy_values_distances}
\end{center}
\end{figure}

\begin{figure}[htb!]
\begin{center}
\includegraphics[width=1.0\linewidth]{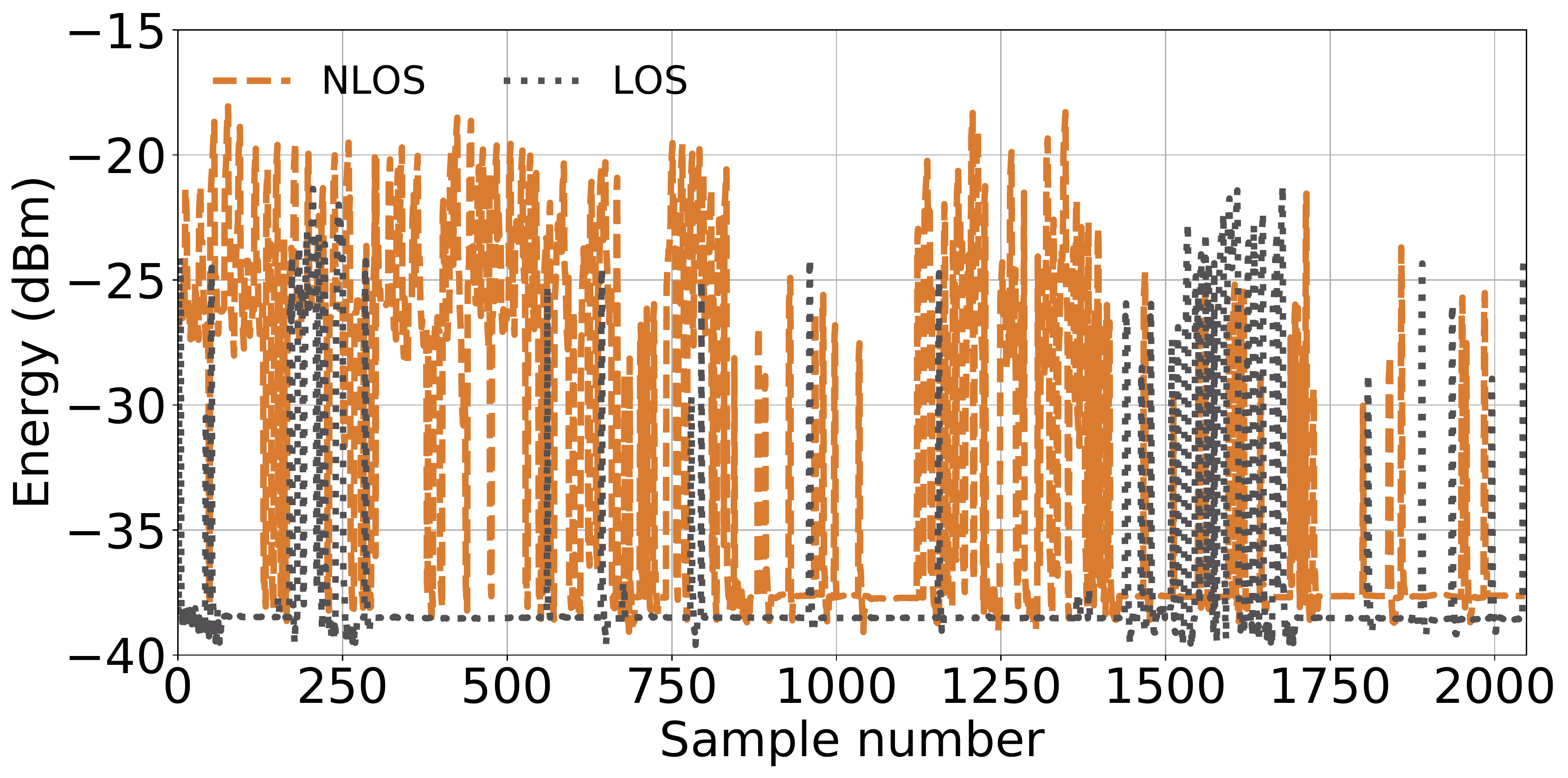}
\caption{\textbf{NLOS vs LOS.} The values of the energy (in dBm) captured for 2048 samples in \mbox{LTE-U} BS while there are 2 \mbox{Wi-Fi} APs at 6 Feet, in NLOS and LOS scenarios. The fewer obstructions, the higher energy is captured.}
\label{fig:energy_values_nlos_los}
\end{center}
\end{figure}

We present the energy of the signals captured in different configurations: (1) Fig.~\ref{fig:energy_values_number_of_wifis} shows the values of energy captured for different number of \mbox{Wi-Fi} APs  (one, two and three), (2) Fig.~\ref{fig:energy_values_distances} demonstrates the scenario with different distances of \mbox{Wi-Fi} APs from the LTE-U, and (3) Fig.~\ref{fig:energy_values_nlos_los} gives insight into energy of the signal in NLOS and LOS scenarios.

We consider in detail the signal from about 1500th sample to 2000th sample in Fig.~\ref{fig:energy_values_number_of_wifis}. It is challenging to distinguish between two or three Wi-Fis \footnote{The Energy values for 4 and 5 Wi-Fi APs are more dense and challenging. In order to better visualize we plotted only 1, 2 and 3 \mbox{Wi-Fi} APs}. The visual inspection of the signals suggests that width of the time-series chunk should be longer than 500 samples. Signals with width of 384 achieve test accuracy below 99\% and signals with width 512 can be trained to obtain 99.68\% of test accuracy. Based on the experiments in Figs. \ref{fig:test-accuracy-chunk-size} and  \ref{fig:energy_values_number_of_wifis}, we find that the best trade-off between accuracy and inference time is achieved for chunk~of~size~512.

\subsection{Transitions between classes}\label{sec:classTransition} 

When we switch to another class (change the state of the system in terms of the number of Wi-Fis), we account for the transition period. If in a given window of 1 second a new \mbox{Wi-Fi} is added, the samples from this first second with new \mbox{Wi-Fi} (or without one of the existing Wi-Fis - when it is removed), the chunk is containing values from $n$ and $n+1$ (or $n-1$) number of Wi-Fis. An easy workaround for the \textit{contaminated} chunk is to change the state of the system to new number of Wi-Fis only after the same class is returned by the model in two consecutive inferences (classifications).

\subsection{Real-time inference}

\begin{figure}[htb!]
\begin{center}
\includegraphics[width=8cm, height = 4cm]{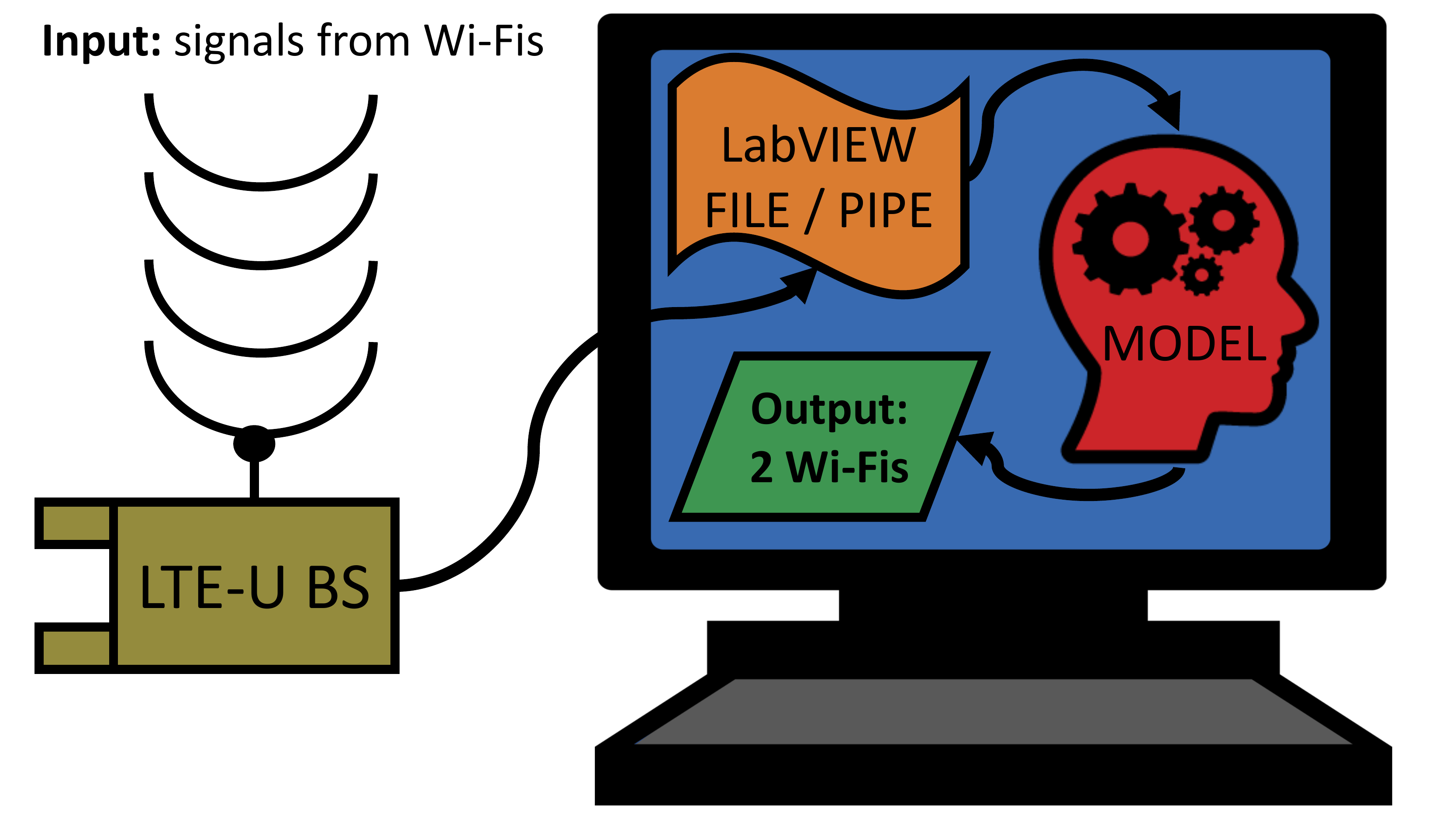}
\caption{The schema of the inference process, where the input received by the \mbox{LTE-U} BS is signals from \mbox{Wi-Fi}s and the output is the predicted number of \mbox{Wi-Fi}s.}
\label{fig:inference}
\end{center}
\end{figure}

We deploy the model in real-time, which is similar to the energy data collection experiment setup, and is shown in Fig.~\ref{fig:inference}. We prepare the model only for the inference task in the following steps. Python scripts load and deploy the trained PyTorch model. We set up the \mbox{Wi-Fi} devices and generate some network load for each device. The \mbox{LTE-U} BS is connected to a computer with the hardware requirements of at least 8 GB RAM (Installed Memory), 64-bit operating system, x64-based processor, Intel(R) Core i7, CPU clock 2.60GHz. The energy of the \mbox{Wi-Fi} transmission signal in a given moment in time is capture using NI LabVIEW. From the program, we generate an output file or write the data to a pipe. The ML model reads the new values from the file until it reaches the time-series chunk length. Next, the chunk is normalized and passed through the model that gives a categorical output that indicates the predicted number of \mbox{Wi-Fi}s in the real-time environment.
 
 \section{Performance comparison between HD, ED, AC and ML methods}\label{comp}
We analyze and study the performance differences between HD, ED, AC and ML methods  for different configuration setups and discus the inference delay. In ML method, we validate the performance on ML real-time inference data. For the final evaluation, we train a single Machine Learning model that is based on the FCN network and used for all the following experiments. The model is trained on the whole dataset of size 3.4 GB, where the train and test sets are of the same size of about 1.7 GB.
\begin{figure}[htb!]
\begin{center}
\includegraphics[width=\linewidth]{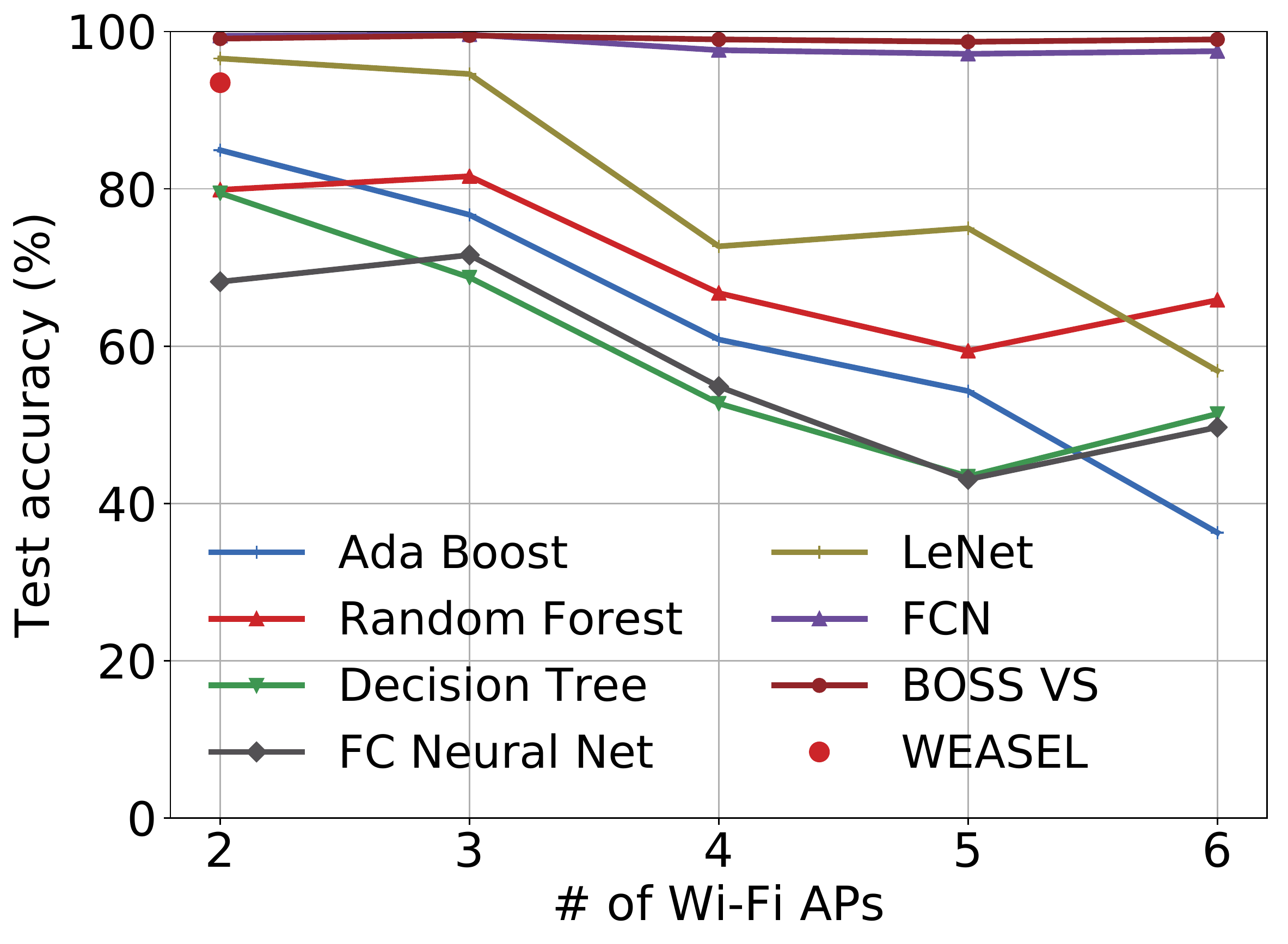}
\caption{Comparison of test accuracy for different ML methods. Number of \mbox{Wi-Fi} APs equals to 2 denotes the Case D configuration (NLOS, 6 feet). Thus, 2 on the x axis corresponds to distinguishing between 1 and 2 \mbox{Wi-Fi} APs, whereas 3 denotes distinguishing between 0, 1, or 2 \mbox{Wi-Fi} APs. Similarly, the values on the x axis (4,5) denote distinguishing from 0 to (x-1) WiFi APs.}
\label{fig:ml-methods-compare}
\end{center}
\end{figure}
\subsection{Comparison between ML methods}

We present comparison between ML methods in Fig.~\ref{fig:ml-methods-compare}. The time-series specific neural network models, such as FCN (\ref{sec:fcn-neural-nets}) as well as BOSS VS (\ref{sec:boss-vs}), perform much better than the general purpose models from scikit-learn library (described in section~\ref{sec:sklearn-ml}). The middle-ground between the two options is a simple two-layer convolutional network called \textit{LeNet}. The main benefit of using FCN (MEDIUM) or BOSS VS is greater model learning capacity than \textit{LeNet} or scikit-learn models. 
There is a negligible difference in terms of test accuracy between the FCN and BOSS VS models. However, the FCN models can scale to much bigger data sizes and we observe that the BOSS VS model often goes out of memory for more than a few GBs of input data. 
Thus, we select FCN as our main Machine Learning (ML) model for all the remaining experiments.

\subsection{Successful Detection at Fixed Distance}
 
\begin{figure}[htb!]
\begin{center}
\includegraphics[width=9cm, height = 5.4cm]{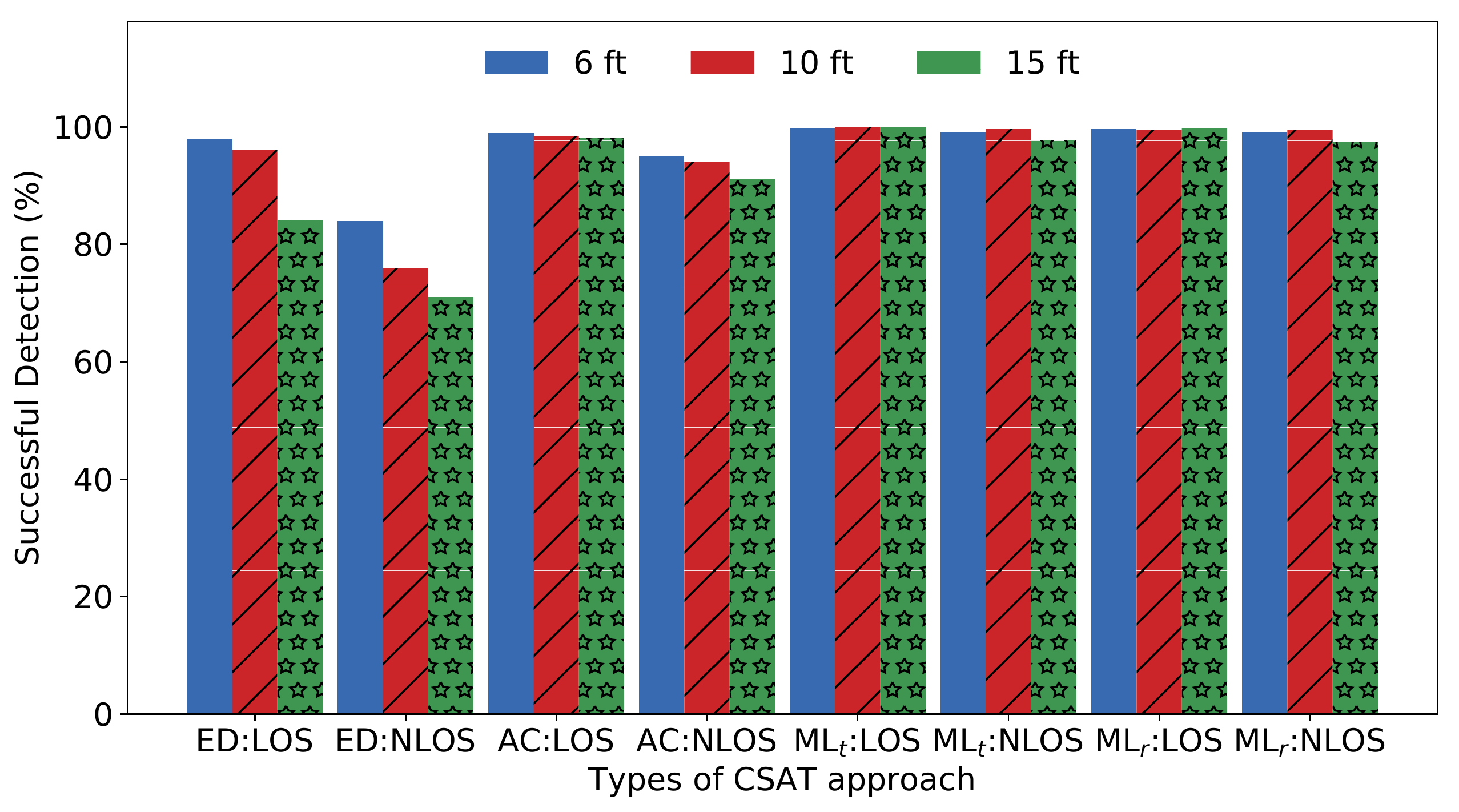}
\caption{Comparison of results for successful detection between ED, AC and ML methods. ML results are presented for the test data (denoted as $ML_t$:) and for the real time inference (denoted as $ML_r$:).}
\label{sample1}
\end{center}
\end{figure}

\begin{table*}
\centering
\caption{Performance of detection for fixed distance configuration setup.}
\begin{tabular}{|*{15}{c|}}  
\hline
\multicolumn{1}{|c}{\cellcolor{Gray} Configuration} &
\multicolumn{1}{|c}{\cellcolor{Gray} Classes} & 
\multicolumn{2}{|c}{\cellcolor{Gray} HD (\%)} & \multicolumn{2}{|c}{\cellcolor{Gray} ED (\%)} & \multicolumn{2}{|c|}{\cellcolor{Gray} AC (\%)} & \multicolumn{2}{|c|}{\cellcolor{Gray} ML (\%)}   \\ \hline
\cellcolor{Gray} Distance &
\cellcolor{Gray} \# of Wi-Fis & \cellcolor{Gray} LOS & \cellcolor{Gray} NLOS & \cellcolor{Gray} LOS & \cellcolor{Gray} NLOS & \cellcolor{Gray} LOS & \cellcolor{Gray} NLOS & \cellcolor{Gray} LOS & \cellcolor{Gray} NLOS \\ \hline
\multirow{5}{*}{6F} & 2 & 100 & 100 & 96 & 91 & 98 & 96 & 98.60 & 99.10 \\ \cline{2-10}
& 3 & 100 & 100 & 88 & 85 & 95 & 90 & 99.10 & 99.50 \\ \cline{2-10}
& 4 & 100 & 100 & 80 & 74 & 87 & 81 & 99.40 & 99.00 \\ \cline{2-10}
& 5 & 100 & 100 & 74 & 62 & 76 & 65 & 99.20 & 98.70 \\ \cline{2-10}
& 6 & 100 & 100 & 62 & 51 & 70 & 59 & 99.30 & 99.0 \\ \hline
\multirow{5}{*}{10F} & 2 & 100 & 100 & 94 & 89 & 97 & 94 & 99.80 & 99.98 \\ \cline{2-10}
& 3 & 100 & 100 & 86 & 82 & 91 & 88 & 99.80 & 99.98 \\ \cline{2-10}
& 4 & 100 & 100 & 78 & 72 & 85 & 79 & 99.80 & 99.90 \\ \cline{2-10}
& 5 & 100 & 100 & 72 & 60 & 75 & 63 & 99.50 & 99.85 \\ \cline{2-10}
& 6 & 100 & 100 & 64 & 54 & 68 & 57 & 99.80 & 99.84 \\ \hline
\multirow{5}{*}{15F} & 2 & 100 & 100 & 92 & 87 & 95 & 90 & 99.80 & 99.80  \\ \cline{2-10}
& 3 & 100 & 100 & 84 & 80 & 85 & 81 & 99.90 & 99.60  \\ \cline{2-10}
& 4 & 100 & 100 & 75 & 70 & 79 & 71 & 99.90 & 99.60  \\ \cline{2-10}
& 5 & 100 & 100 & 70 & 58 & 71 & 64 & 99.60 & 99.50  \\ \cline{2-10}
& 6 & 100 & 100 & 63 & 53 & 66 & 55 & 99.50 & 99.40  \\ \hline
\end{tabular}
\label{mll1}
\end{table*}
We compare the ML performance with HD\footnote{The successful Wi-Fi detection in HD for LOS and NLOS scenario is 100\%. Hence we have not included in the Fig.~\ref{exp}.}, ED and AC approaches using the NI USRP platform as shown in Fig.~\ref{exp}. Similarly we compare the performance of HD by analyzing the Wi-Fi BSSID through wireshark capture.
In the experiment, \mbox{Wi-Fi} APs are transmitting full
buffer data, along with beacon and probe response frames
following the 802.11 CSMA specification. We performed different experiments with 6ft, 10ft and 15ft for LOS and NLOS scenarios.  Fig~\ref{sample1} shows the performance of detection for LOS and NLOS scenarios. In ED and AC based approach the proposed detection algorithm achieves the successful detection on average at 93\% and 95\% for LOS scenario. Similarly, the algorithm achieves 80\% and 90\% for the NLOS scenario. In this work, we show that ML approach can achieve close to 100\% successful detection rate for both LOS and NLOS, and different distance scenarios (6ft, 10ft \& 15ft). We observe the ML approach works close to the performance of HD.

Table~\ref{mll1} shows the performance of detection for fixed distance configuration setup. From, this table the number of Wi-Fis columns represents the number of Wi-Fi APs deployed in the coexistence setup. The number of Wi-Fi AP 2 corresponds to distinguishing between 1 and 2 Wi-Fi APs, whereas 3 denotes distinguishing between  0,  1,  or  2  Wi-Fi  APs and so on. In all cases the performance of ML is close to 100\%. 


\subsection{Successful Detection at Different Configurations}
We verify how the detection works in different configurations. We placed more than two \mbox{Wi-Fi} APs on the same side of the LTE-U BS, unlike the above configuration (i.e., 6ft, 10ft and 15ft) where they were on opposite sides. \mbox{Wi-Fi} AP 1, \mbox{Wi-Fi} AP 2, \mbox{Wi-Fi} AP 3, \mbox{Wi-Fi} AP 4 and \mbox{Wi-Fi} AP 5 are placed at distances of 6 feet, 10 feet and 15 feet from the LTE-U BS respectively. We measured the performance of detection with LOS and NLOS configurations. The goal in this section is to observe the performance of detection in the ML compared with HD, ED, and AC. Some of the possible cases are listed below.
\begin{itemize}
     \item \textbf{Case A:} Only the \mbox{Wi-Fi} AP 1 at 6 feet is ON.
     \item \textbf{Case B:} Only the \mbox{Wi-Fi} AP 2 at 10 feet is ON.
     \item \textbf{Case C:} Only the \mbox{Wi-Fi} AP 3 at 15 feet is ON.
     \item \textbf{Case D:} \mbox{Wi-Fi} AP 1 at 6 feet is ON and \mbox{Wi-Fi} AP 2 at 6 feet is ON.
     \item \textbf{Case E:} The \mbox{Wi-Fi} AP 1 at 6 feet and \mbox{Wi-Fi} AP 3 at 15 feet is ON.
     \item \textbf{Case F:} The \mbox{Wi-Fi} AP 1 at 10 feet and \mbox{Wi-Fi} AP 3 at 15 feet is ON.
     \item \textbf{Case G:} \mbox{Wi-Fi} AP 1 and  \mbox{Wi-Fi} AP 2 at 6 feet is ON and \mbox{Wi-Fi} AP 3 at 15 feet is ON.
   \item \textbf{Case H:} \mbox{Wi-Fi} AP 1 at 6 feet is ON,  \mbox{Wi-Fi} AP 2 at 10 feet is ON, and \mbox{Wi-Fi} AP 3 at 15 feet is ON.
    \item \textbf{Case I:} \mbox{Wi-Fi} AP 1 and  \mbox{Wi-Fi} AP 2 at 6 feet is ON, \mbox{Wi-Fi} AP 3 at 10 feet is ON and \mbox{Wi-Fi} AP 4 at 15 feet is ON.
    \item \textbf{Case J:} \mbox{Wi-Fi} AP 1 at 6 feet is ON,   \mbox{Wi-Fi} AP 2 and \mbox{Wi-Fi} AP 3 at 10 feet is ON and \mbox{Wi-Fi} AP 4 at 15 feet is ON.
    \item \textbf{Case K:} \mbox{Wi-Fi} AP 1 and  \mbox{Wi-Fi} AP 2 at 6 feet is ON, \mbox{Wi-Fi} AP 3 and \mbox{Wi-Fi} AP 4 at 10 feet is ON and \mbox{Wi-Fi} AP 5 at 15 feet is ON.
\item \textbf{Case L:} \mbox{Wi-Fi} AP 1 at 6 feet is ON, \mbox{Wi-Fi} AP 2 , \mbox{Wi-Fi} AP 3 and \mbox{Wi-Fi} AP 4 at 10 feet is ON and \mbox{Wi-Fi} AP 5 at 15 feet is ON.
\item \textbf{Case M:} \mbox{Wi-Fi} AP 1 at 6 feet is ON, \mbox{Wi-Fi} AP 2 at 10 feet is ON, \mbox{Wi-Fi} AP 3, \mbox{Wi-Fi} AP 4 and \mbox{Wi-Fi} AP 5 at 15 feet is ON.
 \item \textbf{Case N:}  \mbox{Wi-Fi} AP 1 and \mbox{Wi-Fi} AP 2 at 6 feet is ON, \mbox{Wi-Fi} AP 3 at 10 feet is ON, \mbox{Wi-Fi} AP 4 and \mbox{Wi-Fi} AP 5 at 15 feet is ON.

\end{itemize}
The different configurations are for LTE-U when it coexists with different number of Wi-Fi APs (from 1 to 5). Table~\ref{t1} shows the better performance for ED and AC compared to the table~\ref{t12}. This is due to fewer number of Wi-Fi AP deployments from Case A to G compared to Case H to N. Hence, the the ED and AC methods can detect the number of Wi-Fi APs close to 80\% for ED and up to 90\% for AC. As the number of Wi-Fi APs increases from 3 to 5 Wi-Fi APs (\emph{i.e.,} Case H to N), we observe substantial degradation in ED performance (to 56\%) and AC performance (to 63\%). Tables~\ref{t1} and ~\ref{t12} show that there is no such degradation in the performance of ML as compared to ED and AC. Hence, we believe that the ML approach is the preferred method for a \mbox{LTE-U} BS in a dense environment to detect the number of \mbox{Wi-Fi} APs and scale back the duty cycle efficiently.

\begin{table*}
\centering
\caption{Performance of detection for different configuration setup (from case A to G).}
\begin{tabular}{|*{18}{c|}}  
\hline
\multicolumn{1}{|c}{\cellcolor{Gray} CSAT Types} &  \multicolumn{2}{|c}{\cellcolor{Gray} CASE A (\%)} & \multicolumn{2}{|c|}{\cellcolor{Gray} CASE B (\%)} & \multicolumn{2}{|c|}{\cellcolor{Gray} CASE C (\%)} & \multicolumn{2}{|c|}{\cellcolor{Gray} CASE D (\%)} & \multicolumn{2}{|c|}{\cellcolor{Gray} CASE E (\%)}  & \multicolumn{2}{|c|}{\cellcolor{Gray} CASE F (\%)} &
\multicolumn{2}{|c|}{\cellcolor{Gray} CASE G (\%)}\\ \hline 
& \cellcolor{Gray} LOS & \cellcolor{Gray} NLOS & \cellcolor{Gray} LOS & \cellcolor{Gray} NLOS & \cellcolor{Gray} LOS & \cellcolor{Gray} NLOS & \cellcolor{Gray} LOS & \cellcolor{Gray} NLOS & \cellcolor{Gray} LOS & \cellcolor{Gray} NLOS &
\cellcolor{Gray} LOS & \cellcolor{Gray} NLOS & \cellcolor{Gray} LOS & \cellcolor{Gray} NLOS \\ \hline
HD & 100 & 100 & 100 & 100 & 100 & 100 & 100 & 100 & 100 & 100 & 100 & 100 & 100 & 100\\ \hline
ED  & 91 & 82 & 90 & 79 & 85 & 78 & 82 & 77 & 80 & 74 & 81 & 72 & 80 & 69\\ \hline
AC & 95 & 91 & 94 & 91 & 92 & 90 & 91 & 90 & 88 & 85 & 88 & 83 & 86 & 77\\ \hline
ML  & 98.80 & 97.96 & 99.94 & 99.37 & 99.96 & 97.74 & 99.46 & 97.80 & 99.21 & 99.14 & 99.32 &  99.10 & 99.56 & 98.44 \\ \hline
\end{tabular}
\label{t1}
\end{table*}

\begin{table*}
\centering
\caption{Performance of detection for different configuration setup (from case H to N).}
\begin{tabular}{|*{18}{c|}}  
\hline
\multicolumn{1}{|c}{\cellcolor{Gray} CSAT Types} &
\multicolumn{2}{|c|}{\cellcolor{Gray} CASE H (\%)} & \multicolumn{2}{|c}{\cellcolor{Gray} CASE I (\%)} & \multicolumn{2}{|c}{\cellcolor{Gray} CASE J (\%)} & \multicolumn{2}{|c|}{\cellcolor{Gray} CASE K (\%)} & \multicolumn{2}{|c|}{\cellcolor{Gray} CASE L (\%)} &
\multicolumn{2}{|c|}{\cellcolor{Gray} CASE M (\%)} &
\multicolumn{2}{|c|}{\cellcolor{Gray} CASE N (\%)} \\ \hline 
& \cellcolor{Gray} LOS & \cellcolor{Gray} NLOS & \cellcolor{Gray} LOS & \cellcolor{Gray} NLOS & \cellcolor{Gray} LOS & \cellcolor{Gray} NLOS & \cellcolor{Gray} LOS & \cellcolor{Gray} NLOS & \cellcolor{Gray} LOS & \cellcolor{Gray} NLOS & \cellcolor{Gray} LOS & \cellcolor{Gray} NLOS & \cellcolor{Gray} LOS & \cellcolor{Gray} NLOS \\ \hline
HD  & 100 & 100 & 100 & 100 & 100 & 100 & 100 & 100  & 100 & 100  & 100 & 100 & 100 & 100 \\ \hline
ED & 80 & 69 & 77 & 67 & 76 & 65 & 68 & 61  & 67 & 56  & 68 & 57 & 66 & 52\\ \hline
AC  & 84 & 74 & 79 & 68 & 77 & 67 & 75 & 65 & 74 & 64 & 72 & 63 & 71 & 59\\ \hline
ML  & 98.70 & 97.36 & 99.24 & 98.26 & 99.76 & 98.24 & 98.83 & 98.06 & 99.11 & 99.04 & 99.02 &  98.05 & 99.96 & 97.74 \\ \hline
\end{tabular}
\label{t12}
\end{table*}

\subsection{Additional Delay to Detect the \mbox{Wi-Fi} AP}
To study the additional delay to detect a Wi-FI AP, we consider a 5 Wi-Fi AP deployment scenario, where, Wi-Fi AP 1 and Wi-Fi AP 2 at 6 feet are ON, Wi-Fi AP 3 and Wi-Fi AP 4 at 10 feet are ON and Wi-Fi AP 5 at 15 feet is ON. We observe a large number of Wi-Fi packets on the air and moreover the LTE-U ON cycle interference impacts the delay in Wi-Fi transmissions. In HD, the total time for the LTE-U BS to decode the BSSID is 1.4 seconds (i.e., \mbox{Wi-Fi} 1st BSSID beacon packet + LTE-U detects $K$ beacon + Additional layer complexity + NI USRP RIO hardware  processing  time). In ED, the  total  time  for  the  energy  based  CSAT algorithm to adopt or change the duty cycle from 50\% to 33\% is 5.9 seconds (i.e., \mbox{Wi-Fi}  1st  beacon  transmission  time + LTE-U detects $K$ beacon (or) data packets time + NI USRP RIO hardware  processing  time) as shown in Table~\ref{t23}. In AC, the total time for the AC based CSAT algorithm to change the duty cycle from 50\% to 33\% is 4.8 seconds (i.e., \mbox{Wi-Fi} 1st L-STF packet frame + LTE-U detects L-STF frame time + NI USRP RIO hardware processing time). In ML, the total time for the CSAT algorithm to adopt the duty cycle from 50\% to 33\% is about 3.1 seconds. This approach is dependent on the chunk size (in this case set to 512). 

\begin{table}
\caption{Other
additional delay to detect the \mbox{Wi-Fi} AP due to the NI hardware 
}
\centering	
\begin{tabular}{|p{3.5cm}| p{2cm}| }
\hline
\cellcolor{Gray} \textbf{CSAT Types} & \cellcolor{Gray} \textbf{NI HW Delay}  \\ 
\hline
Header Decoding (HD) & 1.4 S  \\
\hline
Energy Detection (ED) & 5.9 S  \\
\hline
Auto-correlation (AC) & 4.8 S \\
\hline
Machine Learning (ML) & 3.1 S  \\
\hline
\end{tabular}
\label{t23}
\end{table}

\subsection{FFT compression}

\begin{figure}[htb!]
\begin{center}
\includegraphics[width=\linewidth]{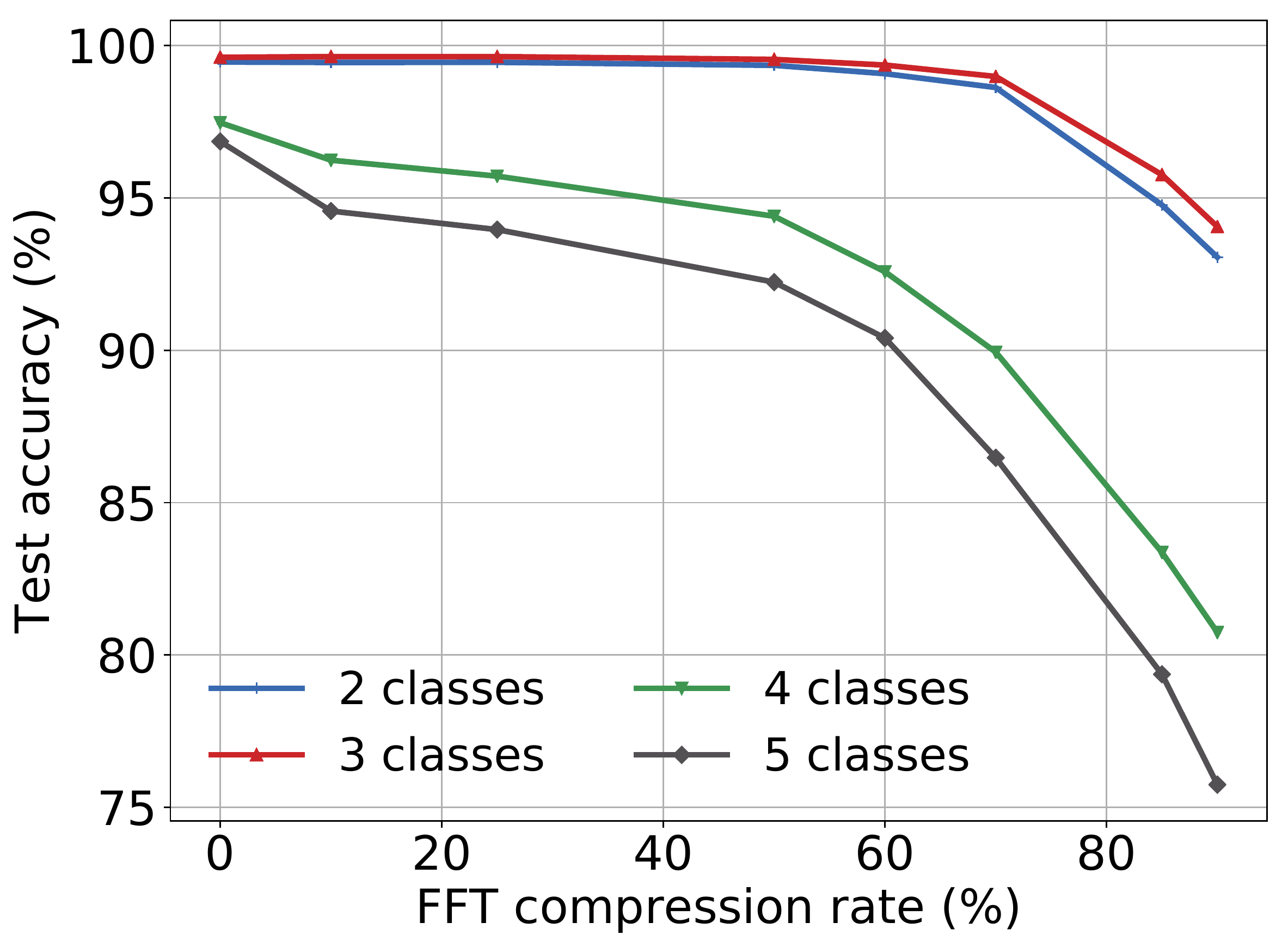}
\caption{Effect of FFT compression embedded into the convolutional layers of the FCN model on test accuracy. We use the Case D configuration for 2 classes and the same configuration with NLOS and 6 feet for the remaining classes.}
\label{fig:wifi-fft}
\end{center}
\end{figure}

We test the FCN model using the FFT based convolutional layers with compression~\cite{dziedzic2019band}. The results are presented in Fig.~\ref{fig:wifi-fft}. We observe that for 2 and 3 classes the data is highly compressible and we can allow up to even 60\% compression with the test accuracy preserved on the level of above 99\%. As we increase the number of classes, the accuracy of the model gracefully degrades and the 60\% compression rate allows us to retain the test accuracy of about 90\% for 5 classes. 

We do not observe a significant difference between the cases with 2 and 3 classes. For 2 classes, we have 1 or 2 \mbox{Wi-Fi} APs and for 3 classes, we distinguish between 0, 1, or 2 \mbox{Wi-Fi} APs. The signal for no \mbox{Wi-Fi} APs is very different and hence easier to classify, than for the remaining signals with active  \mbox{Wi-Fi} APs.

\section{Conclusions and Future Work}\label{sec:conclusion}
We have presented a comprehensive experimental study of different kinds of ML algorithms that could be used to address the problem of identifying the number of active Wi-Fi APs on the air to aid in setting the LTE-U duty cycle appropriately. Additionally, we have compared the performance of the optimum ML algorithm to conventional methods using energy detection and auto-correlation detection and demonstrated superior performance in multiple configurations. We believe that this is the first result that demonstrates the feasibility of using ML on energy values in real-time, instead of packet decoding \cite{chai2016lte}, to reliably distinguish between the presence of different number of \mbox{Wi-Fi} APs. Such a result can have applications beyond LTE-U duty-cycle adaptation, for example in better Wi-Fi frequency management.

We aim to extend this work in the future by distinguishing between LTE-LAA BS and \mbox{Wi-Fi} APs for the coexistence scenario between Wi-Fi, LTE-U and LTE-LAA, thus enabling even finer duty cycle adjustments of a LTE-U BS and improved coexistence with \mbox{Wi-Fi}. Also, we are interested in developing a ML framework that predicts the type of Wi-Fi traffic \emph{i.,e.,} voice, video, or data which in turn can further ensure fair access to the unlicensed spectrum since each traffic-type requires different transmission opportunity times (TXOPs) and per-traffic fairness is more important than per node (Wi-Fi AP) fairness. Similar concepts can also be applied to LTE-LAA/Wi-Fi coexistence deployments and future NR-U/Wi-Fi coexistence in the 6 GHz band.

\section*{ACKNOWLEDGEMENT}\label{p4}
This material is based on work supported by the National Science Foundation (NSF) under Grant No. CNS - 1618920. Adam Dziedzic is supported by the Center For Unstoppable Computing (CERES) at the University of Chicago.

\bibliographystyle{unsrt}
\bibliography{ref}

\end{document}